%% file: main.tex
\documentclass[journal]{IEEEtran}
\IEEEoverridecommandlockouts
\usepackage{amssymb}
\usepackage{amsmath}
\usepackage{url}
\usepackage{pdfpages}
\usepackage[linesnumbered, ruled,vlined]{algorithm2e}
\usepackage[color=green]{todonotes}
\usepackage{multirow}
\usepackage{caption}
\usepackage[utf8]{inputenc}
\usepackage{booktabs}
\usepackage{balance}
\usepackage{pgfplots}
\usepackage{graphicx}
\usepackage{caption}
\usepackage{subcaption}
\usepackage{pgfplots, pgfplotstable}
\pgfplotsset{compat=1.14}
\usetikzlibrary{pgfplots.groupplots,shapes,decorations}

\usepackage{mathtools}

\newlength\figureheight
\newlength\figurewidth
\newlength\smallfigureheight
\newlength\smallfigurewidth
\newlength\largefigureheight
\newlength\largefigurewidth
\usepackage{xcolor}
\usepackage{algorithm2e}
\usepackage{xcolor}
\usepackage{mathtools}
\usepackage{amssymb}
\usepackage{amsmath}
\usepackage{algpseudocode}
\newcommand{\xvbox}[2]{\makebox[#1][l]{#2}}

\SetAlFnt{\footnotesize}

\SetAlCapSty{xAlCapSty}

\SetCommentSty{xCommentSty}

\SetNlSty{mynlfont}{}{} 

\LinesNumbered

\DontPrintSemicolon

\RestyleAlgo{algoruled}

\algtext*{EndIf}% Remove "end if" text

\usepackage{enumitem}
\setenumerate{itemsep=2pt,topsep=2pt,leftmargin=17pt,parsep=0pt,partopsep=0pt}
\setitemize{itemsep=2pt,topsep=2pt,leftmargin=17pt,parsep=0pt,partopsep=0pt}

\newcommand{\paragraphX}[1]{\vskip 4pt \noindent \textit{#1} \hskip .05in}

\makeatletter
\renewcommand\paragraph{\@startsection{paragraph}{4}{0mm} % name, level, indent 
{0.4\baselineskip} % beforeskip: how much space you want before \paragraph 
{-0.6em} % afterskip: how much space you want after \paragraph 
{\normalfont\bfseries%   the font family, etc. you want for \paragraph 
}}% 
\makeatother

\usepackage{scalerel}
\usepackage{tikz}
\usetikzlibrary{svg.path}

\definecolor{orcidlogocol}{HTML}{A6CE39}
\tikzset{
  orcidlogo/.pic={
    \fill[orcidlogocol] svg{M256,128c0,70.7-57.3,128-128,128C57.3,256,0,198.7,0,128C0,57.3,57.3,0,128,0C198.7,0,256,57.3,256,128z};
    \fill[white] svg{M86.3,186.2H70.9V79.1h15.4v48.4V186.2z}
                 svg{M108.9,79.1h41.6c39.6,0,57,28.3,57,53.6c0,27.5-21.5,53.6-56.8,53.6h-41.8V79.1z M124.3,172.4h24.5c34.9,0,42.9-26.5,42.9-39.7c0-21.5-13.7-39.7-43.7-39.7h-23.7V172.4z}
                 svg{M88.7,56.8c0,5.5-4.5,10.1-10.1,10.1c-5.6,0-10.1-4.6-10.1-10.1c0-5.6,4.5-10.1,10.1-10.1C84.2,46.7,88.7,51.3,88.7,56.8z};
  }
}

\newcommand\orcidicon[1]{\href{https://orcid.org/#1}{\mbox{\scalerel*{
\begin{tikzpicture}[yscale=-1,transform shape]
\pic{orcidlogo};
\end{tikzpicture}
}{|}}}}

\usepackage[hidelinks]{hyperref} %<--- Load after everything else
\usepackage{hyperref}
\hypersetup{
    colorlinks=true,
    linkcolor=blue,
    citecolor=green,
    filecolor=magenta,      
    urlcolor=magenta,
}

\newcommand{\system}{{\em Mockingbird}}
\begin{document}

\title{\system: Defending Against Deep-Learning-Based Website Fingerprinting Attacks with Adversarial Traces}

%%% 13 double-column pages (including all) %%%

% \author{Ali Al-Obaidi \href{https://orcid.org/0000-0000-0000-0000}{\includegraphics[scale=1]{figures/orcid_16x16.png}}}

\author{
\IEEEauthorblockN{Mohammad Saidur Rahman\orcidicon{0000-0001-6673-171X}}\IEEEauthorrefmark{4}\IEEEauthorrefmark{2}, 
\IEEEauthorblockA{
%Global Cybersecurity Institute, RIT.\
%\textit{Rochester Institute of Technology},
%Rochester, NY, USA \\
\texttt{\small saidur.rahman@mail.rit.edu}}\\
\IEEEauthorblockN{Mohsen Imani\orcidicon{0000-0002-7934-0301}}\IEEEauthorrefmark{4},
\IEEEauthorblockA{Anomali Inc.,
%Arlington, TX, USA,
\texttt{\small imani.moh@gmail.com}}\\
\and
\IEEEauthorblockN{Nate Mathews\orcidicon{0000-0001-6186-7001}}\IEEEauthorrefmark{2}, 
\IEEEauthorblockA{
%Global Cybersecurity Institute, RIT.\ 
%Rochester, NY, USA \\
\texttt{\small nate.mathews@mail.rit.edu}}\\
\and
\IEEEauthorblockN{Matthew Wright\orcidicon{0000-0002-8489-6347}}\IEEEauthorrefmark{2}, 
\IEEEauthorblockA{
%Global Cybersecurity Institute, RIT.\ 
%Rochester, NY, USA \\
\texttt{\small matthew.wright@rit.edu}}\\
\IEEEauthorrefmark{4} Authors contributed equally\\
\IEEEauthorrefmark{2}Global Cybersecurity Institute, Rochester Institute of Technology, Rochester, NY, USA.
}

\maketitle

% Abstract
\input{0_Abstract.tex}


%%%%%%%% Potential Issue %%%%%%
%% Latency Overhead 
%%%%%%%%%%%%%%%%%%%%%%%%%%%%%%%

% Introduction
\input{1_Intro_Motivation.tex}

% Threat Model
\input{2_Threat_Model.tex}

% Background & Related Work
\input{3_back_rel_work.tex}

% Adv Examples as WF Defense
\input{4_AdvExDef.tex}

% Evaluation
\input{5_Evaluation.tex}

% Discussion
\input{6_discussion.tex}
\input{7_Con_Future.tex}

\paragraphX{\textbf{Resources.}}
The code and datasets of this paper are available at: \textcolor{blue}{\url{https://github.com/msrocean/mockingbird/}}.

% Acknowledgement
\input{8_Acks.tex}

%\paragraphX{\textbf{Resources.}}The code and data will be released upon the acceptance of this paper.
\balance

%{\normalsize \bibliographystyle{acm}
%\bibliography{refs}}

\bibliographystyle{IEEEtran}
\bibliography{refs}

\appendices

%\label{appen_a}
%\input{appen_attack_Scenerios.tex}

%%% SUPPLEMENTARY MATERIAL

\input{appen_A.tex}

\input{appen_D.tex}

\end{document}

%% file: 0_Abstract.tex
\begin{abstract}
Website Fingerprinting (WF) is a type of traffic analysis attack that enables a local passive eavesdropper to infer the victim's activity, even when the traffic is protected by a VPN or an anonymity system like Tor. Leveraging a deep-learning classifier, a WF attacker can gain over 98\% accuracy on Tor traffic. %Existing WF defenses are either very expensive in terms of bandwidth and latency overheads (e.g. two-to-three times as large or slow) or ineffective against the latest attacks. 
In this paper, we explore a novel defense, \system, based on the idea of {\em adversarial examples} that have been shown to undermine machine-learning classifiers in other domains. %To make the technique more robust than existing algorithms for finding adversarial examples,
Since the attacker gets to design and train his attack classifier based on the defense, we first demonstrate that at a straightforward technique for generating adversarial-example based traces fails to protect against an attacker using \textit{adversarial training} for robust classification. We then propose \system, a technique for generating traces that resists adversarial training by moving randomly in the space of viable traces and not following more predictable gradients.
%using randomly selected targets
%and depends on the detector to only determine when the modified trace will be misclassified with high confidence.
%\system~adds padding to a traffic trace in a manner that fools the classifier into classifying it as coming from a different site. 
The technique drops the accuracy of the state-of-the-art attack hardened with adversarial training from 98\% to 42-58\% while incurring only 58\% bandwidth overhead. The attack accuracy is generally lower than state-of-the-art defenses, and much lower when considering Top-2 accuracy, while incurring lower bandwidth overheads. %In addition, the information leakage of \system~is at most 1.9 bits, whereas it is at most 2.2 bits for Walkie-Talkie (W-T) and 2.6 bits for WTF-PAD.
%\saidur{added this last line}
%For most of the cases we examine, the state-of-the-art attack's accuracies against \system~is at least 29\% lower than that of WTF-PAD. Except one case, the attack accuracy of our defense is lower than that of Walkie-Talkie (W-T) defense. The Top-2 accuracy of \system~is at most 56.9\%, while it is 89.7\% for WTF-PAD and 97\% for W-T. In addition, for the most cases, \system's bandwidth overhead is at least 7\% and 15\% lower than that of WTF-PAD and W-T, respectively, showing its promise as a possible WF defense for Tor.
\end{abstract}

%% file: 1_Intro_Motivation.tex
\section{Introduction}

Deep learning has had tremendous success in solving complex problems such as image recognition~\cite{krizhevsky2012imagenet}, speech recognition~\cite{hinton2012deep}, and object tracking~\cite{geiger2012we}. Deep learning models are 
%Nevertheless, it remains a fundamental challenge to ensure that DL models are robust 
vulnerable, however, to \emph{adversarial examples} -- inputs carefully crafted to fool the model~\cite{szegedy2013intriguing}.
%,papernot2017practical,deepfool,carlini2017towards, athalye2018obfuscated}.
Despite a large body of research attempting to overcome this issue, no methods have been found to reliably classify these inputs. In fact, researchers have found that adversarial examples are another side of the coin of how deep learning models are so successful in the first place~\cite{ilyas2019adversarial}. 
%These inputs, known as \emph{adversarial examples}, can confuse models in various settings, such as 

%an effective defense against the underlying vulnerabilities of a DL model, popularly known as {\em adversarial examples}, is still an unsolved problem~\cite{szegedy2013intriguing,papernot2017practical,deepfool,carlini2017towards, athalye2018obfuscated}. These vulnerabilities can be exploited in such way that will cause a trained DL model to behave otherwise.

In this paper, we investigate whether the exploitability of deep learning models can be used for \emph{good}, defending against an attacker who uses deep learning to subvert privacy protections. 
%where the inference of a DL model acts as an attack. 
In particular, we seek to undermine an attacker using deep learning to perform
%want to attack the attacker's trained DL model crafting adversarial examples as a defender. The application domain we are investigating is
{\em Website Fingerprinting} (WF) attacks on the Tor anonymity system. 

%The Tor anonymity system is vulnerable to traffic analysis attacks. One such attack is {\em Website Fingerprinting} (WF), which enables an eavesdropper between the client and the first Tor node on her path to identify which websites the client is visiting.

\begin{figure}[!t]
\centering
	\includegraphics[scale=0.25]{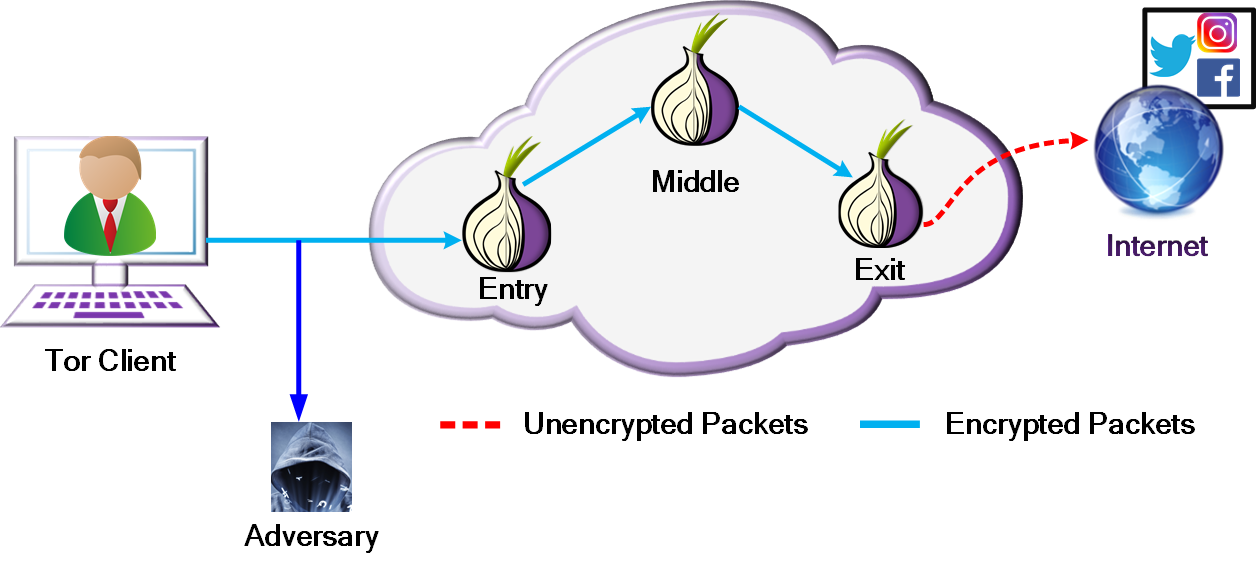}
    \caption{Website Fingerprinting Attack Model.}
    \label{fig:wf-diagram}
    \vspace{-0.2cm}
\end{figure}

WF is a class of traffic analysis attack that enables an eavesdropper between the client and the first Tor node on her path to identify which websites the client is visiting. Figure~\ref{fig:wf-diagram} shows the WF attack model. This local passive adversary could be sniffing the client's wireless connection, have compromised her cable/DSL modem, or gotten access to the client's ISP or workplace network.

The WF attack can be modeled as a supervised classification problem, in which the website domain names are labels and each traffic trace is an instance to be classified or used for training. Recently proposed WF attacks~\cite{rahman2019tik,sirinam2019triplet,bhat2019var, sirinamDF, Rimmer2018} have used deep learning classifiers to great success because of the superior inference capability of deep learning models over traditional machine learning models. The state-of-the-art WF attacks, Deep Fingerprinting (DF)~\cite{sirinamDF} and Var-CNN~\cite{bhat2019var}, utilize convolutional neural networks (CNN) to identify patterns in traffic data. These attacks can achieve above 98\% accuracy to identify sites using undefended traffic in a {\em closed-world} setting~\cite{sirinamDF,bhat2019var}, and both attacks achieve high precision and recall in the more realistic {\em open-world} setting.
%\footnote{WF attacks are evaluated in either the closed-world or open-world settings. The closed-world setting makes assumptions that benefit the attacker, whereas the open-world setting relaxes these assumptions and is more indicative of real-world performance.}

In response to the threat of WF attacks,
%~\cite{sirinam2019triplet,rahman2019tik,sirinamDF,Rimmer2018,Abe2016,kfp,cumul,Wang2014}, 
% Wright2009,Luo2011a,Perry2011,Nithyanand2014,Cai2014wpes,imaniAadvTraces
numerous defenses have been proposed~\cite{glove,Cai2014wpes,Cai2014ccs}. WF defenses perturb the traffic so as to hide patterns and confound the classifier. While some defenses have unacceptably high overheads, two relatively lightweight defenses for Tor have recently been proposed: WTF-PAD~\cite{wtf-pad} and Walkie-Talkie (W-T)~\cite{Wang2017}. State-of-the-art DL attacks, however, have proven effective against both of them~\cite{rahman2019tik,sirinamDF,bhat2019var} and illustrate the need for defenses that can withstand improvements in the attacker's capabilities.
%WTF-PAD offers lower overheads and is considered to be a leading candidate for deployment in Tor~\cite{perry2015}.
%Recent work has shown, however, that WTF-PAD is ineffective against state-of-the-art attacks that achieve over 90\% accuracy on WTF-PAD-defended traffic in a closed-world test~\cite{rahman2019tik,sirinamDF,bhat2019var}. Furthermore, the DF attack is able to reliably narrow down site identities to two candidates when pitted against the W-T defense, achieving a Top-2 accuracy of 98\%~\cite{sirinamDF}. These findings indicate that an effective WF defense can become obsolete in the future against a more advanced classifier.

%These issues have motivated the need for new defenses that perform well against these state-of-the-art DL based attacks. 
This motivates us to investigate a WF defense that can be effective not only against current DL-based attacks but also against possible attack methods that we can foresee.
%We ask ourselves {\em ``how can we perturb the network traffic that will successfully cause misclassification in a current and future DL based WF classifier?"}. Our quest has lead to the investigation of {\em adversarial examples} (AE) that has three intriguing properties:
Adversarial examples are a natural method to turn to for confusing a DL model, so we explore how to create adversarial examples for network traffic.
We find that adversarial examples have three attributes that are valuable for defending against WF: i) {\em high misclassification rates},
ii) {\em small perturbations}, which ensure a low overhead, and iii) {\em transferability}. The transferability property~\cite{szegedy2013intriguing}, in which some adversarial examples can be made to reliably work on multiple models~\cite{carlini2017towards,papernot2017practical}, makes it possible to defend against unknown attacks and potentially even ones with more advanced capabilities.
%Transferability property enables an AE generated with a neural network (NN) remain effective to cause misclassification in a different NN. This particular property will enable a WF defense to be effective not only against current attacks but also against any future advanced attacks.

In this paper, we introduce \system,\footnote{The Northern mockingbird %imperfectly but effectively 
imitates the calls of a wide range of other birds, and one of its call types is known as a {\em chatburst} that it uses for territorial defense:
\url{https://en.wikipedia.org/wiki/Northern_mockingbird}}
a defense strategy using adversarial examples for network traffic traces, which we call \emph{adversarial traces}.
%generated by a DL network. %There are various techniques to generate these examples in the context of image classification~\cite{szegedy2013intriguing,jsma,papernot2017practical,deepfool,carlini2017towards}, however most of them perform a search in the space of possible images. Starting with the original image, the search  moves gradually towards a slightly modified image that is classified differently from the source. This search is typically guided by the gradient of the neural network's loss function or the {\em logits}, the inputs to the neural network's final layer for classification decision. Adversarial examples generated by these methods have had tremendous success in fooling DL models to misclassify with as much as 100\% effectiveness and low amounts of distortion in the images~\cite{Carlini2017AdversarialEA}. This makes adversarial examples intriguing for defending against WF attacks using deep learning, as they could fool the classifier without adding large amounts of distortion in the form of padding or delay.
%While neural networks can be made more robust against adversarial examples, most of these techniques are not effective upon careful analysis~\cite{athalye2018obfuscated,carlini2017magnet} or do not generalize to new approaches to generating the examples~\cite{raghunathan2018certified}.
In the WF context, we cannot simply perform a straightforward mapping of adversarial examples to network traffic. As it is the reverse of when the adversary is applying the adversarial examples, the WF attacker gets to know what the WF defender is doing to generate the examples. In particular, he can use the defense as implemented in open-source Tor code to generate his own adversarial traces and then use them to train a more robust classifier. This {\em adversarial training} approach has been shown to be effective when the classifier knows how the adversarial traces are being generated~\cite{adv_train_ensemble}.

To address this, we propose a novel technique to generate adversarial traces that seeks to limit the effectiveness of adversarial training. In particular, we increase the randomness of the search process and reduce the influence of the design and training of the targeted deep learning model in finding a good adversarial example. To this end, as we search for the new trace, we select a random target trace and gradually reduce the distance from the modified trace to the target. We also change to other randomly selected targets multiple times during the search. The deep learning model is only used to give a confidence value on whether the current trace fools the classifier, and we do not access the loss function, the logits, or any other aspect of the model. The resulting adversarial traces can go in many different directions from the original source traces instead of consistently following the same paths that result from, e.g., following the gradient of the loss function. In this way, the technique selects paths that are hard to find through adversarial training. Further, each new trace generated from the same source typically ends near a different target each time, helping to reduce the attacker's Top-$k$ accuracy.

Extensive evaluation 
%in both white-box\footnote{The attacker's and the defender's NN is the same NN.} and black-box\footnote{The attacker's and the defender's NN are two different NN.} setting 
shows that \system\footnote{This work is an extended version of a short paper~\cite{rahmanAdvtrace2018}.}~reliably causes misclassification in a deep learning classifier hardened with adversarial training using moderate amounts of bandwidth overhead. Our results hold even when the attacker uses a significantly more powerful classifier than the target classifier used by \system~to produce the adversarial examples.
%With this technique, \system~reliably causes misclassification in a deep-learning classifier hardened with adversarial training using only moderate amounts of bandwidth overhead. 

\paragraphX{\bf{\textit{Contributions.}}} In summary, the key contributions of this work are as follows:
%\vspace{0.5cm}
\begin{itemize}
    \item We propose \system, the first WF defense to leverage the concept of adversarial examples.
    
    \item We show how algorithms for generating adversarial examples in computer vision fail as a defense in the WF setting, motivating more robust techniques.
    
%    \item White-box evaluation shows that \system~ significantly reduces accuracy of the state-of-the-art WF attack hardened with adversarial training from 98\% to 29\%-55\%, depending on the scenario. The bandwidth overhead is 56\% for full-duplex traffic, which is better than both W-T and WTF-PAD.
    
    \item Our evaluation shows that \system~significantly reduces accuracy of the state-of-the-art WF attacks hardened with adversarial training from 98\% to 38\%-58\% attack accuracy, depending on the scenario. The bandwidth overhead is 58\% for full-duplex traffic, which is better than both W-T and WTF-PAD.
    
    \item We show that \system~makes it difficult for an attacker to narrow the user's possible sites to a small set. The best attack can get at most 72\% Top-2 accuracy against \system, while its Top-2 accuracy on W-T and WTF-PAD is 97\% and 95\%, respectively.

    \item Using the WeFDE framework~\cite{infoleak}, we measure the information leakage of \system, and find that it has less leakage for many types of features than either W-T or WTF-PAD.
\end{itemize}

Our investigation of this approach provides a promising first step towards leveraging adversarial examples to undermine WF attacks and protect user privacy online.

% mkw -- this kind of thing is good as a way to make reviewers happy
%        that you're addressing their requested changes. But we don't
%        need to prime them to be unhappy with these issues right off
%        the bat.
%We note that \system~ has significant implementation challenges, perhaps most importantly that finding suitable traces requires significant computational time. At a minimum, \system~ has potential for use in server-side WF defenses~\cite{Cherubin2017}, in which the computation need only be done once for each significant change to the site. Additionally, this work serves as a potential first step towards more practical general-purpose defenses that build on our findings. We do discuss the promising directions to overcome the challenges and we encourage the community to investigate further in this research direction. 

%% file: 2_Threat_Model.tex
\section{Threat \& Defense Model}
\label{models}

%In this section, we discuss about the website fingerprinting (WF) threat model and the defense model.

\subsection{Threat Model}\label{attack-model}
We assume that the client browses the Internet through the Tor network to hide her activities (see Figure~\ref{fig:wf-diagram}). The adversary of interest is {\em local}, which means the attacker is positioned somewhere in the network between the client and Tor guard node. The attacker is assumed to already know the identity of the client. His goal is to detect the websites that the client is visiting. A {\em local} adversary can be an eavesdropper on the user’s local network, local system administrators, Internet service provider, any networks between the user and the entry node, or the operator of the entry node. 
%Another assumption of this adversary is that he is {\em passive} meaning that the adversary only taps the traffic between the client and the entry node (as shown in Figure~\ref{fig:wf-diagram}), he just observes and records the traffic traces, he does not have the capacity to drop, delay, or modify the real packets, or inject fingerprints in the traffic traces. 
The attacker is {\em passive}, meaning that the he only observes and records the traffic traces that pass through the network. He does not have the ability to drop, delay, or modify real packets in the traffic stream.

In a website fingerprinting (WF) attack, the attacker feeds the collected network traffic into a trained machine-learning or deep-learning classifier. For the purpose of training, the WF attacker first needs to collect traffic of various sites by operating a Tor client. Since is not feasible to collect traffic for all the sites on the web, %To tackle this issue, an attacker collects traffic of different sites in which he wants to keep an eye on and based on his interest. This set of websites is known as monitored sites. The sites which are not in the monitored sites are known as unmonitored sites.
%To address this, 
the attacker identifies a set of {\em monitored} sites that he wants to track. The attacker limits the scope of his attack to the identification of any website visits that are within the monitored set. The set of all other sites is known as the {\em unmonitored} set.

WF attacks and defenses are evaluated in two different settings: {\em closed-world} and {\em open-world}. In the closed-world setting, we assume that the client is limited to visiting only the {\em monitored} sites. The training and testing set used by the attacker only include samples from the monitored set. The closed-world scenario models an ideal setting for the attacker and is not indicative of the attack's real-world performance. From the perspective of developing a WF defense, demonstrating the ability to prevent closed-world attacks is thus sufficient to show its effectiveness.

In contrast, the open-world scenario models a more realistic setting in which the client may visit websites from both the monitored and unmonitored sites. In this setting, the attacker trains on the monitored sites and a representative (but not comprehensive) sample of unmonitored sites. The open-world classifier is then evaluated against both monitored and unmonitored sites, where the set of unmonitored sites used for testing does not intersect with the training set.%Over time the performance of the attacks has been improved such that they could reach over 98\% accuracy rate in the closed-world scenario and their false positive rate is less than 0.4\% in the open-world setting.

\subsection{Defense Model}
\label{defense_model}
%\todo[inline]{NM: to saidur-I did some major revision to this section, make sure I didn't change the intended meaning}
The purpose of a WF defense is to prevent an attacker who observes a traffic trace from determining accurately to which site the trace belongs. 
To achieve this, the real traffic stream must be manipulated in some way. Because traffic is bidirectional, the deployment of a successful WF defense requires participation from both the client and a cooperating node in the Tor circuit. 
%It is difficult, if not impossible, for the client to defend their traffic alone as they have control over only outgoing traffic in the communication stream. To defend both directions in the network stream, the client must have the cooperation of one of the nodes in its Tor circuit. 
We call this node the {\em bridge}.\footnote{Tor bridges are usually used for evading censorship, but they can be used for prototyping WF defenses such as used in WTF-PAD~\cite{wtf-pad}.} To defend against eavesdroppers  performing WF attacks, the {\em bridge} could be any node located between the adversary and the client's destination server, making it so that the adversary only has access to the obfuscated traffic stream. Since the guard node knows the IP address of the client and can thus act as as a WF adversary, it is better to set up the bridge at the middle node, which cannot directly identify the client. %In order to reduce additional bandwidth requirements on the full Tor circuit, it is usually ideal that the {\em bridge} be located on the guard node. %The defense we propose assumes the cooperating node to be the guard node, however it is possible to deploy our defense on any secured node in the Tor circuit.
%WF defenses require two parties in the network to control the both sides of the traffic, upload traffic and download traffic. We assume the part of the defense controlling the upload traffic is located in the client side. The part of the defense controlling the download side of the traffic can be deployed on a {\em bridge}. The {\em bridge} can be located somewhere between the adversary and the webservers. The possible options for deploying the {\em bridge} can be the guard node or webservers. The purpose of WF defense is to make the traffic traces indistinguishable among different sites.

%% file: 3_back_rel_work.tex
\section{Background \& Related Work}
\label{backandrelated}

%In this section, we review the prior works on WF attacks and defenses in Tor.

\subsection{WF Attacks}\label{wf-attack-related-work}
%The study of website fingerprinting dates back to the 1990s when the security researchers first investigated information leaks in encrypted HTTP requests that could expose the identity of the site being visited~\cite{WagnerSchneier, And98trafficanalysis}. 
%The threat of website fingerprinting against Tor was first evaluated by Herrmann et al.~\cite{herrmann2009website} in 2009. Their attack used a Naive Bayes classifier with a feature set created from the distribution of the frequencies of packet lengths. However this reliance on packet length lead to the attack's failure against Tor, with only 3\% accuracy. Tor sends data in fixed 512-byte units, known as {\em cells}, which makes packet-length features ineffective. %Since then, researchers have incrementally improved the performance of WF attacks through the development of better feature sets and the application of new classification algorithms.
%In this section we explore several notable attacks that utilize hand-crafted features as well as recent deep-learning based attacks. %In recent years, three such attacks using manually designed features have emerged as benchmarks for WF attacks, and we briefly discuss each in turn.

Website fingerprinting attacks have applied a variety of classifiers. The three best attacks based on manual feature engineering are $k$-NN~\cite{Wang2014}, CUMUL~\cite{cumul}, and $k$-FP~\cite{kfp}, which all reached over 90\% accuracy in closed-world tests on datasets with 100 samples per site. In the rest of this section, we examine the more recent deep-learning-based WF attacks.

\paragraphX{SDAE.} 
%Recently, deep-learning techniques have attracted the attention of privacy researchers due to their excellent performance in image recognition tasks. 
The first to investigate using deep-learning techniques for WF were Abe and Goto~\cite{Abe2016}, who developed an attack based on Stacked Denoising Autoencoders (SDAE). Their model was trained on raw packet direction, represented by a sequence of ``+1" and``-1" values for outgoing and incoming packets, respectively. Despite this innovation, their attack achieved a lower accuracy rate than the previous state-of-the-art attacks at only 88\% in the closed-world setting. %The reason for their low accuracy rate would seem to be due to the limited number of samples in the dataset (100 per site), as DL models require more data to train when compared to other classification algorithms. 

\paragraphX{Automated Website Fingerprinting (AWF).} Rimmer et al.~\cite{Rimmer2018} proposed using deep learning to bypass the feature engineering phase of traditional WF attacks. To more effectively utilize DL techniques, they collected a very large dataset of 900 sites with 2,500 trace instances per site. They applied several different DL architectures---SDAE, Convolutional Neural Network (CNN), and Long Short-Term Memory (LSTM)---on the traffic traces. They found that their CNN model outperforms the other DL models they developed, obtaining 96\% accuracy in the closed-world setting. 

\paragraphX{Deep Fingerprinting (DF).} Sirinam et al.~\cite{sirinamDF}  developed a deeper CNN model that reached up to 98\% accuracy rate in the closed-world setting using a dataset of 100 sites with 1,000 instances each. They also examined the effectiveness of their model against WF defenses, where they showed that their model can achieve concerningly high performance against even some defended traffic. Most notably, their attack achieved 90\% accuracy against WTF-PAD~\cite{wtf-pad} and 98\% Top-2 accuracy against Walkie-Talkie~\cite{Wang2017}.

\paragraphX{Var-CNN.} Recently, Bhat et al.~\cite{bhat2019var} developed a more sophisticated WF attack based on the ResNet CNN architecture and attained 98.8\% closed-world accuracy. 

\vskip 0.2cm 
\noindent 
We evaluated \system~on both the DF model and Var-CNN model in the black-box setting.

\subsection{WF Defenses}\label{wf-defeses-bkgrd}
To defeat WF attackers, researchers have explored various defense designs that generate cover traffic to hide the features present in website traffic. WF defenses are able to manipulate the traffic stream with two operations: sending dummy packets and delaying real packets. These manipulations, however, come at a cost: sending dummy packets adds an additional bandwidth overhead to the network, while delaying packets adds latency overhead that directly impacts the time required to load the page. Several studies have thus tried to balance the trade-off between the WF defense's overhead and efficacy of the defense against WF attacks. In this section, we review these WF defenses.

\paragraphX{Constant-rate padding defenses.} This family of defenses transmits traffic at a constant rate in order to normalize trace characteristics. BuFLO~\cite{dyer:sp12} is the first defense of this kind, and it sends the packets in the same constant rate in both directions. The defense ends transmission after the page has finished loading and a minimum amount of time has passed. The overhead of the traffic is governed by both the transmission rate and the minimum time threshold for the stopping condition. Moreover, although the defense covers fine-grained features like burst information, course-grained features like the volume and load time of the page still leak information about the website. Tamaraw~\cite{Cai2014ccs} and CS-BuFLO~\cite{Cai2014wpes} extend the BuFLO design with the goal of addressing these issues. %Since nearly all sites send more traffic to the client than the client sends to them, Tamaraw and CS-BuFLO transmit the download and upload packets at different fixed rates.
To provide better cover traffic, after the page is loaded, Tamaraw keeps padding until the total number of transmitted bytes is a multiple of a fixed parameter. Similarly, CS-BuFLO pads the traffic to a power of two, or to a multiple of the power of the amount of transmitted bytes. All of these defenses are expensive, requiring two to three times as much time as Tor to fetch a typical site and more than 100\% bandwidth overhead.

\paragraphX{Supersequence defenses.} This family of defenses depends on finding a {\em supersequence} for traffic traces. To do this, these defenses first cluster websites into anonymity sets and then find a representative sequence for each cluster, such that it contains all the traffic sequences. All the websites that belong to the same cluster are then molded to the representative supersequence. This family includes Supersequence~\cite{Wang2014}, Glove~\cite{glove}, and Walkie-Talkie~\cite{Wang2017}. Supersequence and Glove use approximation algorithms to estimate the supersequence of a set of sites. The traces are then padded in such a way so as to be equivalent to its supersequence. However, applying the molding directly to the cell sequences creates high bandwidth and latency costs. Walkie-Talkie (WT) differs from the other two defenses in that it uses anonymity sets of just two sites, and traces are represented as burst sequences rather than cell sequences. Even with anonymity sets of sizes of just two, this produces a theoretical maximum accuracy of 50\%. %WT reduces the browser to a {\em half-duplex} mode of operation, in which the browser sends new requests only after the responses to all the previous requests have been received. WT seeks to pair sensitive websites with nonsensitive websites, and due to WT's theoretical maximum attacker accuracy of 50\%, an attacker is unable to reliably determine whether a webpage visit was sensitive or not. %Bursting molding of a particular burst index of a site is done by taking the maximum burst length between that site and a nonsensitive site. 
Wang and Goldberg report just 31\% bandwidth overhead for their defense, but also 34\% latency overhead due to the use of half-duplex communication. Against WT, the DF attack achieved 49.7\% accuracy and 98.4\% \emph{top-2} accuracy, meaning that it could effectively identify the two sites that were molded together but not distinguish between them~\cite{sirinamDF}. 

\paragraphX{WTF-PAD.} Shmatikov and Wang~\cite{ap} proposed \emph{Adaptive Padding (AP)} as a countermeasure against end-to-end traffic analysis. Juarez et al.~\cite{wtf-pad} proposed the WTF-PAD defense as an adaptation of AP to protect Tor traffic against WF attacks. WTF-PAD tries to fill in large delays between packets {\em (inter-packet arrival times)}. Whenever there is a large inter-packet arrival time (where "large" is determined probabilistically), WTF-PAD sends a fake burst of dummy packets. This approach does not add any artificial delays to the traffic. Juarez et al. show that WTF-PAD can drop the accuracy of the $k$-NN attack from 92\% to 17\% with a cost of 60\% bandwidth overhead. Sirinam et al.~\cite{sirinamDF}, however, show that their DF attack can achieve up to 90\% accuracy against WTF-PAD in the closed-world setting. 

\paragraphX{Application-level defenses.} Cherubin et al.~\cite{Cherubin2017} propose the first WF defenses designed to work at the application layer. They proposed two defenses in their work. The first of these defenses, ALPaCA, operates on the webserver of destination websites. %This defense was specifically designed for special websites known as onion sites (further explained in Section \ref{onion-sites}).
%Cherubin et al. argued their defense was particularly suited for Onion Sites, which are typically more interested in preserving user privacy and would thus be more willing to run a defense. 
ALPaCA works by altering the size distribution for each content type, e.g. PNG, HTML, CSS, to match the profile for an average onion site. In the best case, this defense has 41\% latency overhead and 44\% bandwidth overhead and reduces the accuracy of the CUMUL attack from 56\% to 33\%. Their second defense, LLaMA, operates exclusively on the client. It adds random delays to HTTP requests in an effort to affect the order of the packets by manipulating HTTP request and responses patterns. LLaMA drops the accuracy of the CUMUL attack on Onion Sites from 56\% to 34\% at cost of 9\% latency overhead and 7\% bandwidth overhead.

%\input{3_1_back_onion_sites.tex}

%% file: 4_AdvExDef.tex
\section{Preliminaries}\label{adv-wf-defense}

%\subsection{Adversarial Examples \& Adversarial Training}
%\label{adv_examples}

%\subsubsection*{\textbf{Adversarial Examples}}
%Adversarial examples have become a matter of concern since
%\paragraphX{\textbf{Adversarial Examples.}}
\subsection{Adversarial Examples}
Szegedy et al.~\cite{szegedy2013intriguing} were the first to discover that otherwise accurate ML and DL image classification models could be fooled by image inputs with slight perturbations that are largely imperceptible to humans. These perturbed inputs are called {\em adversarial examples}, and they call into question the robustness of many of the advances being made in machine learning.
%. These are inputs that are crafted from the distribution of correctly classified samples, but with slight perturbations that result in a misclassification.
%discovered that several machine learning models including neural networks are vulnerable to the {\em adversarial examples}. 
The state-of-the-art DL models can be fooled into misclassifying adversarial examples with surprisingly high confidence. For example, Papernot et al.~\cite{papernot2017practical} show that adversarial images cause a targeted deep neural network to misclassify 84\% of the time. 
%Adversarial examples are  

The idea of creating adversarial examples is to modify samples from one class to make them be misclassified to another class, where the extent of the modification is limited. More precisely, given an input sample {\em x} and target class {\em t} that is different from actual class of {\em x} ($t \neq {C}^{*}(x)$), the goal is to find ${x}^{'}$ which is close to {\em x} according to some distance metric and $C({x}^{'}) = t$. In this case, ${x}^{'}$  is a {\em targeted} adversarial example since it is misclassified to a particular target label {\em t}. An {\em untargeted} adversarial example, on the other hand, may be misclassified to any other class except the true class (${C}^{*}(x)$).

%The reason behind the existence of adversarial examples is a mystery. There is some speculation about the cause of adversarial examples, such as the extreme non-linearity of deep neural networks, overfitting, insufficient model averaging, and insufficient regularization. Due to the effectiveness of adversarial examples, there is an arms race between defenses against adversarial examples and new types of adversarial example attacks. 

In response to the threat of adversarial examples, many defense techniques have been introduced to make classifiers more robust against being fooled. Recent research~\cite{athalye2018obfuscated,Carlini2017AdversarialEA} shows that almost none of these recent defense techniques are effective. In particular, we can generate adversarial examples that counter these defense techniques by including the defense techniques directly into the optimization algorithm used to create the adversarial examples. We can also overcome many defense approaches by simply increasing the amount of perturbation used~\cite{carlini2017magnet}.

%\paragraphX{\textbf{Properties of AE.}}
\subsection{Properties of Adversarial Examples}
Adversarial examples have three major properties that make them intriguing for us in WF defense: i) robust misclassification, ii) small perturbations, and iii) transferability. We now explain the effect of each of these properties in a WF defense.

%\begin{itemize}
\paragraphX{\textbf{Robust Misclassification.}} An effective defense should be able to fool a trained WF classifier consistently in real-world conditions. Adversarial examples have been shown to work reliably and robustly for images, including for cases in which the viewpoint of the camera cannot be fully predicted, such as fooling face recognition~\cite{sharif2019general} and self-driving cars~\cite{eykholt2018physical}.
    
\paragraphX{\textbf{Small Perturbations.}} To fool the classifier, the defense will add padding packets to the original network traffic. Ideally, a WF defense should be lightweight, meaning that the number of padding packets should be constrained to keep bandwidth consumption low. By using small perturbations to achieve misclassification, an effective WF defense based on adversarial examples can also be lightweight.

\paragraphX{\textbf{Transferability.}} 
Recent research shows that defenses such as WTF-PAD~\cite{wtf-pad} and W-T~\cite{Wang2017}, which defeated the state-of-the-art attacks available at that time, are seriously underminded by the more recent and advanced attacks~\cite{rahman2019tik,sirinamDF}. Given that the attacker could use any classifier for WF, the defense should extend beyond current attacks to other possible attacks as well.
%This finding indicates that we need a WF defense which should be effective not only against current state-of-the-art attacks but also any future and advanced attacks that we are not aware of. Transferability property of AE can provide us that advantage. 
Adversarial examples provide the ability for this due the the \emph{transferability} property, which indicates that they can be designed to attack a given classifier and at least sometimes also fool other classifiers~\cite{carlini2017towards}. Further, there are techniques that work in a \emph{black-box} setting, where the classifier is completely unknown to the attacker~\cite{papernot2017practical,liu_adv_ex}. This property is very important for WF defense, since we cannot predict the attacker's classifier in advance.

%We can evaluate the strength of a WF defense to defeat any future attacks by evaluating the defense in a black-box setting where the attacker adapts a whole different model to attack the defended network traffic. Note that, due to the nature of the earlier proposed WF defense, it was not feasible to predict the performance of a WF defense against any future advanced attacks. However, it is possible for \system~to predict and provide us some confidence about its future performance. We define white-box and black-box attack for a WF defense in Section~\ref{exp_method}.
%\end{itemize}

%%%%%%%%%%%%%%%%%%%%%%%%%%%%
%\subsubsection*{\textbf{Adversarial Training}}
%\paragraphX{\textbf{Adversarial Training.}}
\subsection{Adversarial Training}
\label{adv_training}

%\todo[inline]{Moh: we should explain why we design \system}
%\todo[inline]{MSR: this section should be hrere. We need to talk about adv training before even mention it in the next section.}
Recent research shows that adversarial training increases the robustness of a model by incorporating adversarial examples in the training data~\cite{madry2017towards,adv_train_ensemble}. The idea is to train a network with adversarial examples so that they can be classified correctly. This approach is limited, as it does not adapt well to techniques for generating adversarial examples that haven't been trained on. In the WF setting, however, the classifier has the advantage of knowing how the adversarial examples are being generated, as they would be part of the open-source Tor code. Thus, adversarial training is a significant concern for our system.

\begin{figure}
    \centering
    \includegraphics[scale=0.4]{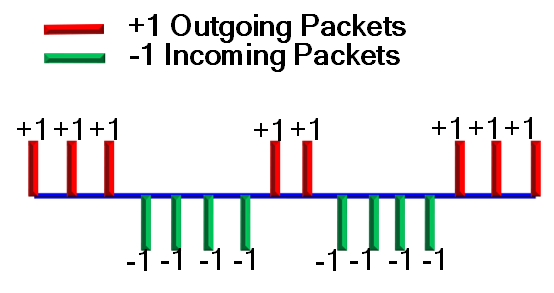}
    \caption{A visual representation of bursts.}
    \label{fig:bursts}
    \vspace{-0.2cm}
\end{figure}

%\paragraphX{\textbf{Data Representation.}}
\subsection{Data Representation}
%To develop a WF defense based on the adversarial examples, first we need to create a representation for the traffic traces to apply the adversarial examples algorithms to them. 
Following the previous work~\cite{wtf-pad,Wang2017,sirinamDF}, we model the traffic trace as a sequence of incoming (server to client) and outgoing (client to server) bursts. The incoming and outgoing packets are represented as $-1$ and $+1$, respectively. We define a burst as a sequence of consecutive packets in the same direction. An example of the burst representation is shown in~Figure~\ref{fig:bursts}. 
% (see Figure~\ref{fig:burst}). 
% mkw -- this doesn't belong here
%The traffic traces themselves can be considered in two communication modes: half-duplex, as used in Walkie-Talkie~\cite{Wang2017}, and full-duplex, which is how Tor currently operates. 
Given this, we can increase the length of any burst by sending padding packets in the direction of the burst.  
%\matt{wait, do we split bursts like this in the defense?}
%\saidur{we split the bursts like fig~2 in the defense}
%Moreover, we can decrease the length of the burst by sending dummy packets in the opposite direction of an ongoing burst to interrupt the ongoing burst 
%\saidur{we don't do this, we increase the bursts by adding packets in the direction of the bursts for both incoming and outgoing, we don't interrupt the burst pattern, this sentence is not correct for our defense and we should remove this line}. 
We cannot, however, decrease the size of the bursts by dropping real packets due to the retransmissions this would cause, changing the traffic patterns and adding delays for the user.

%%%%%%%%%%%%%%%%%%%%%%%%%%%%%%%%%%
%%%%%%%%%%%%%%%%%%%%%%%%%%%%%%%%%%%
\section{\system~ Design}
\label{defense_design}

We now motivate and describe the design of the \system~defense in detail. We start by evaluating the performance given by adapting existing adversarial-example techniques, which are not effective against adversarial training, and then examine the \system~design in detail. %to  the preliminaries about the properties of adversarial examples (AE), including data representation for a WF defense.
%We now motivate and describe the design of the \system~defense, starting with a brief background on adversarial examples and adversarial training, then showing the limitations of applying existing techniques for generating adversarial examples, and finally showing the \system~design in detail.

%%%%%%%%%%%%%%%%%%%%%%%%%%%%
\subsection{Applying Existing Methods in WF Defense}% as a WF Defense}
\label{failed_methods}
%Supplemental Materials

Several different algorithms have been proposed for generating adversarial examples within the field of computer vision, including the Fast Gradient Sign Method (FGSM)~\cite{fsg}, the Iterative Fast Gradient Sign Method (IGSM)~\cite{kurakin2016adversarial}, the Jacobian-Based Saliency Map Attack (JSMA)~\cite{jsma}, and optimization-based methods~\cite{Carlini2017AdversarialEA, liu_adv_ex}. For our initial exploration of adversarial examples in WF defense, we examine the technique proposed by Carlini and Wagner (C\&W)~\cite{carlini2017towards}.

%In 2017, Carlini and Wagner proposed a powerful optimization-based algorithm to generate adversarial examples~\cite{carlini2017towards}.
This method is shown to defeat the defensive distillation approach of blocking adversarial examples~\cite{distillation}. The algorithm is successful with 100\% probability. We modified their technique to suit our needs to generate adversarial traces out of the burst sequences. This algorithm is designed to work on images, which are 2D, so we modified it to work on 1D traffic traces.

%Given a sample $x$ and a model $F$, the algorithm finds a perturbation $\delta$ that makes $x' = x + \delta$ to be misclassified to any other class than $C(x)$ ($C(x) = {argmax}_{i}{F(x)}_{i}$), or in the case of a targeted attack, it is classified to target class $t$. The algorithm tries to find $x'$ that is similar to $x$ based on distance metric $D$. The distance metrics can be an ${L}_{p}$-norm such as ${L}_{0}$, ${L}_{2}$, or ${L}_{\infty}$.  The algorithm is formulated as follows:
%\begin{IEEEeqnarray*}{rCl}
%\textrm{min} \quad {\parallel\delta\parallel }_{p} + c.f(x + \delta)\\
%\textrm{such that} \quad x + \delta \in {[0,1]}^{n}
%\IEEEyesnumber
%\end{IEEEeqnarray*}

%The algorithm will find $\delta$ such that it minimizes the distance metric, which is ${l}_{p}$ norm, and the objective function $f(x + \delta)$. $c$ is a constant to scale both the distance metric and objective function in the same range. Carlini and Wagner~\cite{carlini2017towards} used binary search to find the proper value for $c$. They explored several objective functions and found two taht work the best. For a targeted attack scenarios with target class $t$, the best objective function is:
%
%\begin{IEEEeqnarray*}{rCl}
%f({x'}) = \max_{i \neq t}({F({x'})}_{i}) - {F({x'})}_{t}
%\IEEEyesnumber
%\end{IEEEeqnarray*}
%
%For non-targeted attack scenarios where the true class for sample $x$ is class $y$, the best objective function is:
%\begin{IEEEeqnarray*}{rCl}
%f({x'}) =  {F(x')}_{y} - \max_{i \neq y}({F({x'})}_{i})
%\IEEEyesnumber
%\end{IEEEeqnarray*}

We evaluated the performance of this technique in two different WF attack scenarios: {\em without-adversarial-training} and {\em with-adversarial-training}. The without-adversarial-training scenario represents the scenario most typically seen in the adversarial example literature, in which the classifier has not been trained on any adversarial instances. In this scenario, we generated the adversarial examples against a target model and tested them against the different WF attacks trained on the original traffic traces. We find that our adversarial traces are highly effective against the WF attacks. The accuracy of DF~\cite{sirinamDF} is reduced from 98\% to 3\%, and the accuracy of CUMUL~\cite{cumul} drops from 92\% to 31\%. The adversarial traces generated using this method are highly transferable, as we generated them against a target CNN model similar to the one proposed by Rimmer et al.~\cite{Rimmer2018}, and they are effective against both DF and CUMUL.

Unfortunately, this scenario is not realistic, as it is likely (and usually assumed) that the attacker can discern what type of defense is in effect and train on representative samples. This is represented by the \emph{with-adversarial-training} scenario. In this scenario, the C\&W technique fails completely, with the DF attack reaching 97\% accuracy. In addition, we also investigated a method that combines aspects of our \system~system with C\&W, but this also proved to be ineffective as a WF defense. We discuss the details of these evaluations in the {\em Appendices}.% tuned C\&W in Appendix~\ref{tunedcw}. %Moreoever, we have generated the adversarial traces using FGSM~\cite{fsg} and JSMA~\cite{jsma} and we discovered the same findings in both scenarios.

The results of this evaluation led to a very important insight: the scenario in which the effectiveness of adversarial examples are typically evaluated is notably different than that of a WF defense. In particular, the attacker has the advantage of going second, which means that the classifier can be designed and trained after the technique is deployed in Tor. Thus, techniques that excel at producing adversarial examples for traditional attacks are poorly suited for our problem. In response to this discovery, we focused our efforts on the development of a new technique designed specifically for our needs. We discuss our method in the following section.

\input{4_adv_def_algo.tex}

%% file: 4_adv_def_algo.tex
%\section{Adversarial Traces as WF Defense}
%\label{df-1}

\subsection{Generating Adversarial Traces}
\label{genadvtraces}

\begin{figure}[ht]
    \centering
    \vskip -0.3cm
	\includegraphics[width=\linewidth]{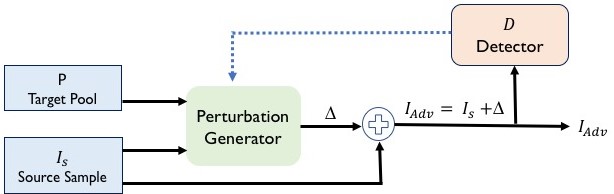}
    \caption{\system~Architecture.}
    \label{fig:mockingbird}
\end{figure}
%\vskip -0.5cm

%\input{4_algo_code}

%Note that the defense that we are proposing is more suited for Onion sites because Onion Sites are more willing to collaborate in deploying the defense on their servers to protect their users. For the sake of evaluation we used the data from the regular web. We believe that the results should be consistent on the Onion sites given that the WF attacks are less effective on Onion sites~\cite{unique-ccs2017}, which make the defenses more effective.

We now introduce {\system}, a novel algorithm to generate adversarial traces that more reliably fool the classifier in the adversarial training setting. The ultimate goal is to generate untargeted adversarial traces that cause the classifier to label a traffic trace as coming from some site other than the original site, i.e. to generate an untargeted sample. We find, however, that it is more effective to generate targeted samples, i.e. to select specific other sites for the current sample to attempt to mimic. Much like its namesake (the bird), \system~uses a variety of targets, without much importance on which target it mimics at any given moment.  
%, i.e. to generate an untargeted example, we find it more efffective to generated targeted examples instead. 
%In practice, this means that we can select from a range of potential targets to move o

To defend a given trace, the {\em source sample}, \system~first generates a set of potential target traces selected randomly from the traces of various sites other than the source site. It then randomly picks one of these traces as the {\em target sample} and gradually changes the source sample to get closer to the target sample. The process stops when a trained classifier called the {\em detector} determines that the class of the sample has changed (see Figure~\ref{fig:mockingbird}). Note that it does not need to have changed to the target sample's class, as the goal is to generate an untargeted adversarial trace. The amount of change applied to the source sample governs the bandwidth overhead of \system, and as such should be minimized. 

Unlike most other algorithms used to generate adversarial examples, \system~does not focus on the loss function (like FGSM and IGSM) or logits (like Carlini \& Wagner) of the detector network. Instead, it aims to move the source sample towards the target sample and only uses the detector network to estimate the confidence with which the trace is misclassified. This lessens the reliance on the shape of the detector network and helps to generate adversarial traces that are more robust against adversarial training. As we demonstrate in the \emph{Supplementary Materials}, using an optimization method to move towards the target sample results in a system that is much less robust.

%We next explain the \system~algorithm in detail.
%In order to limit the amount of change added to the traces, \system~changes the source samples until they are no longer classified as their source classes. \system~uses a trained classifier, called a {\em detector}, to evaluate whether the perturbed source sample will be misclassified. If the {\em detector} identifies the perturbed source sample is still in the source class, \system~keeps changing the source sample toward the target sample until it leaves the source class. In the following section we explain the \system~algorithm in detail.

%\todo[inline]{Moh: below paragraph should be added later}
%This pool can have samples, i.e. traffic traces, from either a closed-world dataset  or a open-world dataset. The closed-world target pool contains samples drawn randomly from the closed-world dataset. Any samples from the source sample's class are excluded. 
%Therefore, the closed-world target pool has samples that may belong to any classes except the source class. 
%Similarly, the open-world target pool contains samples drawn randomly from the open world dataset.

\paragraph*{{\system~{\em Algorithm}}} We assume that we have a set of sensitive sites $\mathcal{S}$ that we want to protect. We train a {\em detector} $f(x)$ on a set of data from $\mathcal{S}$. We discuss the design and training of $f(x)$ in Section~\ref{evaluations-df}. We consider traffic trace ${I}_{s}$ as an instance of source class $s \in \mathcal{S}$. Our goal is to alter ${I}_{s}$ to become $I'_s$ such that it is classified to any other class $t$, $t = f({I'}_{s})$ and $t\neq s$.
%should be trained on only sensitive sites or both sensitive and non-sensitive sites

${I}_{s}$ is a sequence of the bursts, ${I}_{s} = \left [ {b}^{I}_{1},  {b}^{I}_{2},...,{b}^{I}_{n}\right ]$, where $n$ is the number of the bursts. Burst is defined as the sequence of packets in a direction (i.e. incoming or outgoing)~\cite{bhat2019var, sirinamDF, Wang2014}. The length of each burst (i.e. number of packets in each burst), ${b}^{I}_{i}$, makes the sequence of bursts. Usually, the number of packets in each burst vary widely. The only allowed operation on a burst ${b}^{I}_{i}$ is to add some positive values ${\delta}_{i} >= 0$ to that burst, {\em ${b}^{I}_{i} = {b}^{I}_{i} + {\delta}_{i}$}. The reason for using ${\delta}_{i} >= 0$ is that we can only increase the size of a burst. If ${\delta}_{i} < 0$, that would mean we should drop some packets to reduce the size of a burst, and dropping real packets means losing data and requires re-transmission of the dropped packet. 
%\todo{oh, the obvious missing thing is that we can add new bursts. More expensive to compute all the possibilities, but likely valuable.}
%\todo{Moh: yeah, you are right, adding whole fake burst will be good in terms of defense, I already thought about it, but could not find a way to generate them through the algorithm, may be add them randomly in the trace, which is not efficient way}
%\matt{well, we could explore it, but it's not critical right now relative to deployment issues}
To protect source sample ${I}_{s}$, we first select $\tau$ potential target samples from other classes $t_i \neq s$. We then select the target $t$ as the one nearest to the source sample based on the ${l}_{2}$ norm distance\footnote{We also performed some preliminary experiments with {\em Manhattan } distance and found no significant improvement.}. This helps to minimize overhead, as we will move the source towards the target. More formally, we pick a {\em target pool} ${P}_{s}$ of ${p}$ random samples from other classes, ${P}_{s} = \left [ {{I}^{0}}_{0},{{I}^{1}}_{1}, .., {{I}^{p}}_{m} \right]$, where ${{I}^{j}}_{i}$ is the $j$-th sample in the target pool and belongs to target class ${t}_{i} \neq s$. The target sample ${I}_{t}$ is selected as shown in Equation~\ref{eq_dis_target}.
%
%we pick a {\em target pool} ${P}_{{I}_{s}}$ of $p$ random samples from other classes, ${P}_{{I}_{s}} = \left [ {{I}^{0}}_{{t}_{0}},{{I}^{1}}_{{t}_{1}}, .., {{I}^{p}}_{{t}_{m}} \right ]$. ${{I}^{j}}_{t_i}$ is the j-th sample in the target pool and belonging to target class $t_i \neq s$. We want to pick a target sample and re-cast the source sample to be classified as that target class. To decrease amount of change to the source sample, since padding adds bandwidth overhead, we pick the sample from the target pool that is nearest to the source sample. We use the ${l}_{2}$ norm distance to find the nearest target sample to the source sample from the pool (see Equation~\ref{eq_dis_target}). Then we modify the source sample to decrease the distance to this target sample.
\begin{IEEEeqnarray*}{rCl}\label{eq_dis_target}
{I}_{t} = \underset{I \in P_s}{\operatorname{argmin}}  D(I_s , I)
\IEEEyesnumber\\ \label{eq_dis}
D(x,y) = {l}_{2}(x - y)
\IEEEyesnumber
\end{IEEEeqnarray*}
%

%Our goal is to increase the volumes of selected bursts in the source sample such that the source sample is not classified as class $s$ and the amount of change is as small as possible to minimize the bandwidth overhead. 

\input{4_algo_code}

To make the source sample leave the source class, we change it with the minimum amount of perturbation in the direction that makes it closer to the target ($I_t$). We define $\Delta$ as the perturbation vector that we add to the source sample to generate its defended form ${I}^{new}_{s}$. 
\begin{IEEEeqnarray*}{rCl}
\Delta = \left [ {\delta}_{0},  {\delta}_{1},\cdots,{\delta}_{n}\right ]  \quad ({\delta}_{i} >= 0)
\IEEEyesnumber
\end{IEEEeqnarray*}
%\forall i \in [0,\cdots, n]:
\begin{IEEEeqnarray*}{c}
{I}^{new}_{s} = {I}_{s} + \Delta
\IEEEyesnumber
\end{IEEEeqnarray*}

We need to find a $\Delta$ that adds the least amount of perturbation to the source sample while still making it closer to the target sample. Therefore, we find $\Delta$ that minimizes distance $D({I}^{new}_{s}, {I}_{T})$. To do so, we compute the gradient of the {\em distance} with respect to the input. Note that most work in adversarial example generation uses the gradient of the loss function of the discriminator network rather than distance, and this may make those techniques more sensitive to the design and training of the classifier. The gradient points in the direction of steepest ascent, which would maximize the distance. Therefore, we compute the gradient of the negative of the distance with respect to the input, and we modify the source sample in that direction towards the target sample. In particular:
\begin{IEEEeqnarray*}{rCl}
\nabla  (-D(I, {I}_{T})) = - \frac{\partial D(I, {I}_{T})}{\partial I} = \left [ - \frac{\partial D(I, {I}_{T})}{\partial {b}_{i}}  \right ]_{i\in [0, \cdots,n]}
\IEEEyesnumber\\
\end{IEEEeqnarray*}
Where ${b}_{i}$ is the i-th burst in input $I$.

To modify the source sample, we change bursts such that their corresponding values in $\nabla (-D(I, {I}_{T}))$ are positive. Our perturbation vector $\Delta$ is: 
\begin{IEEEeqnarray*}{rCl}
{\delta}_{i}  = \left\{\begin{matrix}
- \alpha \times \frac{\partial D(I, {I}_{T})}{\partial {b}_{i}} &  - \frac{\partial D(I, {I}_{T})}{\partial {b}_{i}} > 0\\ 
$0$ &   - \frac{\partial D(I, {I}_{T})}{\partial {b}_{i}} \leqslant  0
\end{matrix}\right.
\IEEEyesnumber
\end{IEEEeqnarray*}
where $\alpha$ is the parameter that amplifies the output of the gradient. The choice of $\alpha$ has an impact on the convergence and the bandwidth overhead. If we pick a large value for $\alpha$, we will take bigger steps toward the target sample and add more overhead, while small values of $\alpha$ require more iterations to converge. We modify the source sample by summing it with $\Delta$, (${I}^{new}_{s} = {I}_{s} + \Delta$). We iterate this process, computing $\Delta$ for ${I}_{s}$ and updating the source sample at each iteration until we leave the source class, $f({I}^{new}_{s}) \neq s$ or the number of iterations passes the maximum allowed iterations. Note that at the end of each iteration, we update the current source sample with the modified one, $I_s = {I}^{new}_{s}$.  Leaving the source class means that we have less confidence on the source class. So we fix a threshold value, $\tau_{c}$, for measuring the confidence. If the confidence of the {\em detector} on the source class is less than the threshold ($f_s({I}^{new}_{s}) < \tau_{c}$), \system~will stop changing the source sample (${I}_{s}$). 
%\todo[inline]{MOHSEN: we fix a small threshold value (i.e. 0.01) for measuring the confidence. Can you help me explain more?}
%\todo[inline]{MI: Saidur: I think is enough?}

As we only increase the size of the bursts where $- \frac{\partial D(I, {I}_{T})}{\partial {b}_{i}} > 0$, we may run into cases that after some iterations $\nabla  (-D(I, {I}_{T}))$ does not have any positive values or all the positive values are extremely small such that they do not make any significant changes to ${I}_{s}$. In such cases, if ${I}^{new}_{s} - {I}_{s}$ is smaller than a threshold $\tau_{D}$  % (we use different threshold values such as 0.001, 0.0001,  and 0.00001) 
for $\lambda$ consecutive iterations (we used $\lambda=10$), and we are still in the source class, we select a new target. In particular, we effectively restart the algorithm by picking a new pool of potential target samples, selecting the nearest target from the pool, and continuing the process. It is to note that, the source sample at this point is already in the changed form ${I}^{new}_{s}$ and the algorithm starts changing from ${I}^{new}_{s}$. In this process, the confusion added by \system~in the final trace is proportional to the number of targets changed to reach the final adversarial trace. The pseudo code of \system~algorithm is presented in Algorithm~\ref{alg:mbalgo}.

%we drop the previous target samples in the target pool, fill the pool with new target samples, and continue the process.
%\matt{I don't understand -- it doesn't appear to pick a new target from the existing pool? It throws it out and picks a whole new pool? Why}\todo{No, it refills the pool with the new targets, because we pick the nearest target from the pool to the source sample, so, we refill the pool to may have change to find the one which is closer, we can pick from the current pool,too, but I thought refilling gives us better chance to find the closest sample}\matt{then there is no pool! We grab potential targets, pick the closest and throw out the rest right away.}
%\todo[inline]{MOHSEN: this part is also not clear."we refill the pool with new samples and pick a new target sample ${I}_{T}$ to continue the process."} 
%\todo[inline]{MI to SR: is that enough?} 

%\subsubsection*{Limiting Top-N Accuracy}

%% file: 4_algo_code.tex
\begin{algorithm}

% functions
\SetKwFunction{cumprod}{cumprod}
\SetKwFunction{length}{length}
\SetKwFunction{zeros}{zeros}
\SetKwFunction{ceil}{ceil}

% input/ouput names
\SetKwInOut{Input}{Input}
\SetKwInOut{Output}{Output}

% caption
\caption{Generate Adversarial Traces.\label{alg:mbalgo}}
%\system~to 
\Input{%
		\xvbox{2mm}{$\mathcal{S}$} -- set of sensitive sites \\
		\xvbox{2mm}{$\mathcal{D}$} -- detector $f(x)$ \\
		\xvbox{3mm}{$\alpha$} -- amplifier \\
		\xvbox{3mm}{$\delta$} -- iterations before selecting a new target \\
		\xvbox{3mm}{$\tau_{c}$} -- confidence threshold \\
		\xvbox{3mm}{$\tau_{D}$} -- perturbation threshold\\
		\xvbox{3mm}{$N$} -- maximum number of iterations\\		
        \xvbox{3mm}{$p$} -- number of targets to pick for the pool\\
        \xvbox{3mm}{$I_{s}$} -- instance of site $s$ to protect\\ \xvbox{3mm}{$b_{j}$} -- bursts in $I_s$, $j \in [1,\ldots,n]$\\
	  }
\Output{%
		\xvbox{2mm}{$I'_s$} -- altered (adversarial) trace of $I_s$
	   }

  \BlankLine % blank line for spacing
  
  % start of the pseudocode
   \xvbox{4mm}{$Y_s \leftarrow \mathcal{D}(I_s)$ \tcp*{label of $I_s$}}
   
   %   \xvbox{4mm}{$I_sc$} -- copy of the trace $I_s$ \tcp*{for comparing the change}  
   
   \xvbox{24mm}{$P_{s} \leftarrow random(p,S-\{s\})$ \tcp*{target pool}} \BlankLine % blank line for spacing     
   \xvbox{5mm}{${I}_{T} \leftarrow \underset{I \in P_{s}}{\operatorname{argmin}}  D(I_s , I)$ \tcp*{target sample}\;}

   \xvbox{5mm}{$I'_s \leftarrow I_s$}
   
  \For{iter\ in\ $[1,\ldots,N]$}{

    \xvbox{45mm}{$\nabla  (-D(I'_{s}, {I}_{T})) \leftarrow \left [ - \frac{\partial D(I'_S, {I}_{T})}{\partial {b}_{j}}  \right ]_{j\in [1, \cdots,n]}$} %-- gradient of the negative of the distance 
    %\tcp*{gradient of the negative of the distance}

    \BlankLine % blank line for spacing    
    \xvbox{1mm}{$\Delta \leftarrow \alpha \times \nabla  (-D(I'_{s}, {I}_{T})),\ $where$\ - \frac{\partial D(I'_{s}, {I}_{T})}{\partial {b}_{j}} > 0$}
    %\tcp*{amplify by $\alpha$}
    
    \xvbox{2mm}{$I'_{s} \leftarrow I'_{s} + \Delta$}
    \BlankLine % blank line for spacing      
     \tcc*{Compute the label and confidence for $I'_s$}
     \xvbox{24mm}{$Y'_s,\ P(I'_{s}) \leftarrow \mathcal{D}(I'_{s})$} 
     %-- predicted label and prediction score of perturbed $I_s$
    \BlankLine % blank line for spacing     
    %\xvbox{2mm}{$\delta \leftarrow I_{s} - I_{sc}$}
    
     \tcc*{End if the source class confidence is low}
    \If{$Y'_s \neq Y_{s}$ \textbf{and} $P(I'_{s}) < \tau_{c}$}{break\;}
    
    % started from 1 now. 
    % Also, presumably, want to do this on the 10th, 20th, etc. iterations, not the 11th, 21st, etc.
     \tcc*{Pick a new target after $\delta$ iterations}
      \If{$i \bmod \delta = 0$ \textbf{and} $Y'_s = Y_{s}$ \textbf{and} $I'_{s} - I_{s} < \tau_{D}$}{
        $P_{s} \leftarrow random(p,S-\{s\})$\ \tcp*{new target pool}
        %$\nabla P_{samples} \leftarrow random(P_{s})$\ \tcp*{new target pool}
       ${I}_{T} \leftarrow \underset{I \in P_{s}}{\operatorname{argmin}} D(I_s, I)$\ \tcp*{new target sample}}
       %$\nabla {I}_{T} \leftarrow \underset{I \in P_{samples}}{\operatorname{argmin}}  D(I_s , I)$\ 
  }
 % \xvbox{10mm}{$I'_{s} \leftarrow I_{s} $} -- assign perturbed $I_{s}$ to $I'_{s}$
 \xvbox{10mm}{\textbf{return} $I'_{s}$}
\end{algorithm}

%% file: 5_Evaluation.tex
\section{Evaluation}
\label{evaluations-df}

%\paragraphX{\textbf{Defense Evaluation.}} 
%We evaluate the effectiveness of our defense in two settings: i) white-box and ii) black-box. In a white-box setting, the attacker is assumed to have the knowledge of the weights of the defender network and uses the same network to attack the defense. In a black-box setting, the attacker uses any classifier other than defender classifier. We also evaluate a special circumstance where we assume the defender's network is weaker than the attacker's classifier.

\input{5_1_Datasets.tex}

\input{tables/dataset_split.tex}

\subsection{Experimental Method}
\label{exp_method}
%\todo[inline]{MOHSEN: why do we split in this way? what about DF model's train/val/test split}
%\paragraphX{Dataset Split - Attacker \& Defender Set}

\system~needs to train a detector $f(x)$ with instances of a variety of sites to generate adversarial traces. We use both Rimmer et al.'s CNN model~\cite{Rimmer2018}, which we refer to as AWF, and Sirinam et al.'s more powerful DF model~\cite{sirinamDF} for detector models.
%We use the {\em Deep Fingerprinting (DF)} CNN model designed by Sirinam et al.~\cite{sirinamDF} as our detector and 
We train these models on the traces of the Detector Set. Sirinam et al. suggest using an input size of 5,000 packets, but our padded traces have more traffic, so we use an input size of 10,000 packets, which is the 80th percentile of packet sequence lengths in our defended traces. Sample from the Detector Set are {\em only} used to train the detector. The source samples from Adv Set ($I_s \in \mathcal{A} $) are used to generate adversarial traces for the training, validation, and testing of the adversary's classifier. To evaluate the defended traffic, we reform the defended traces from the burst level representation to the direction level representation where ``+1'' and ``-1'' indicate outgoing and incoming packets, respectively, following the previous research.

\begin{figure*}[!t]
\centering
    \begin{subfigure}[h]{0.5\textwidth} \centering
        \includegraphics[scale=0.4]{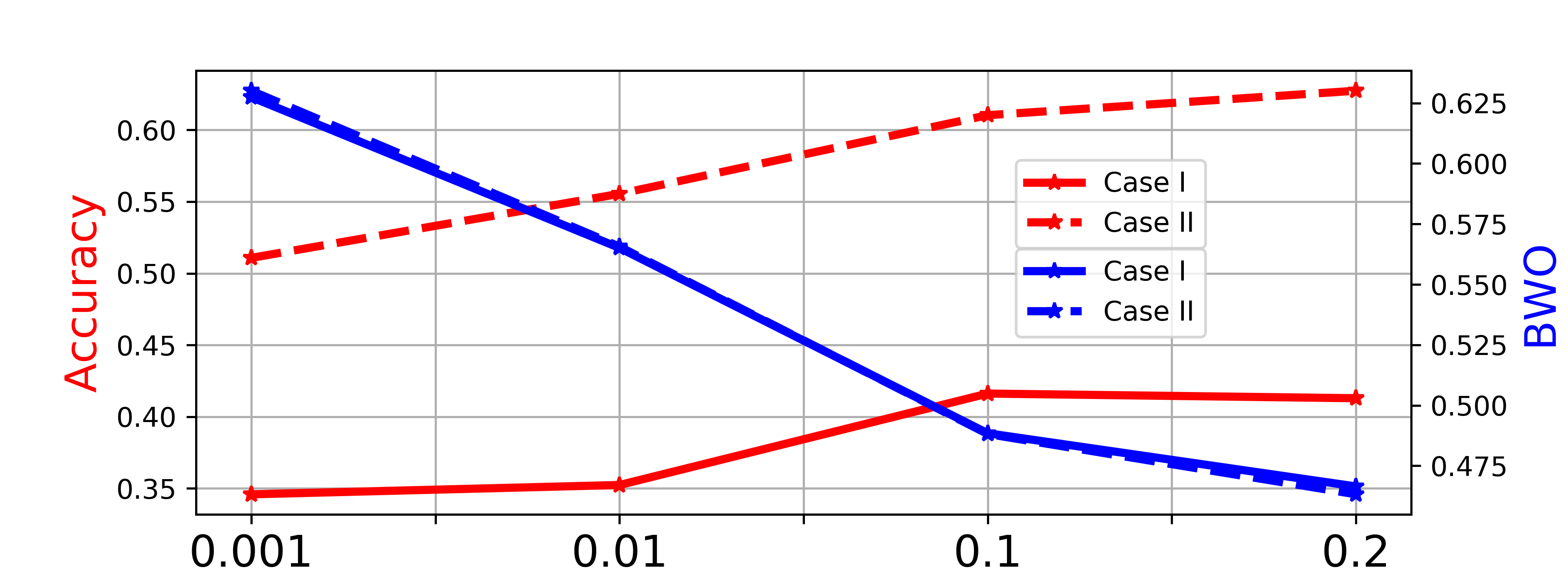}
        \vskip -0.2cm
	\caption{Confidence Threshold Value~($\tau_c$)} % caption line
	\label{fig:prob_threshold_fd}
    \end{subfigure}%
    ~
    \begin{subfigure}[h]{0.5\textwidth} \centering
        \includegraphics[scale=0.4]{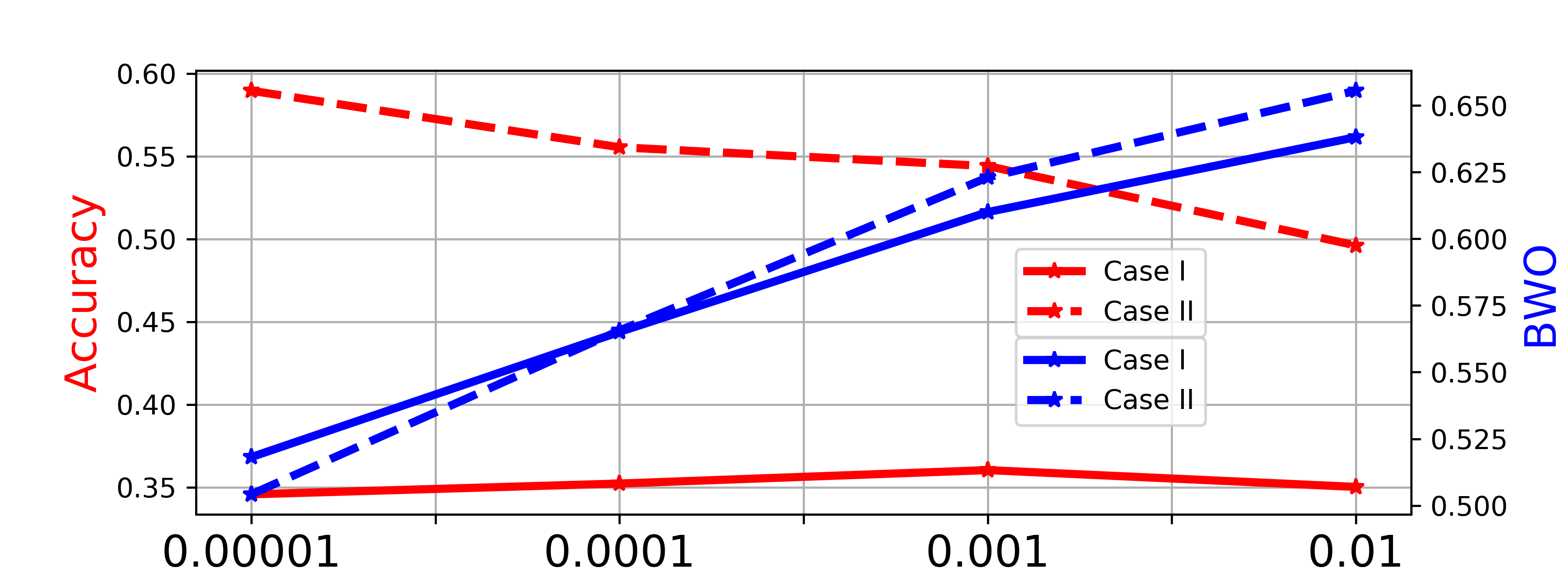}
        \vskip -0.2cm
        \caption{Perturbation Threshold Value ($\tau_D$)}
        \label{fig:change_threshold_fd}
    \end{subfigure}%
\vskip -0.2cm    
\caption{\textit{Full-duplex}: the attacker's accuracy and the bandwidth overhead (BWO) of the generated samples as we vary the probability threshold value (Figure~\ref{fig:prob_threshold_fd}) and perturbation threshold value (Figure~\ref{fig:change_threshold_fd}).}
% Dashed lines show the results of Case I and solid lines show the results of Case II.
\label{fig:threshold_fd}\end{figure*}

\begin{figure*}[!t]
\centering
    \begin{subfigure}[h]{0.3\textwidth} \centering
        \input{figures/alpha.tex}
        \vskip -0.2cm
	\caption{$\alpha$ Value for full-duplex} % caption line
	\label{fig:alpha_full_dup}
    \end{subfigure}%
    ~
    \begin{subfigure}[h]{0.3\textwidth} \centering
        \input{figures/C_1_full_acc_bw.tex}
        \vskip -0.2cm
        \caption{Case I}
        \label{fig:adv-cw-DF-acc}
    \end{subfigure}%
    ~
    \begin{subfigure}[h]{0.3\textwidth} \centering
        \input{figures/C_2_full_acc_bw.tex}
		\vskip -0.2cm
		\caption{Case II}
        \label{fig:adv-open-DF-acc}
    \end{subfigure}%    
\vskip -0.2cm    
\caption{\textit{Full-duplex}: the attacker's accuracy and the bandwidth overhead (BWO) of the generated samples as we vary the $\alpha$ value (Figure~\ref{fig:alpha_full_dup}) and number of iterations (Figure~\ref{fig:adv-cw-DF-acc} and \ref{fig:adv-open-DF-acc}).}
% Dashed lines show the results of Case I and solid lines show the results of Case II.
\label{fig:adv-DF-acc_full_dup}
\end{figure*}
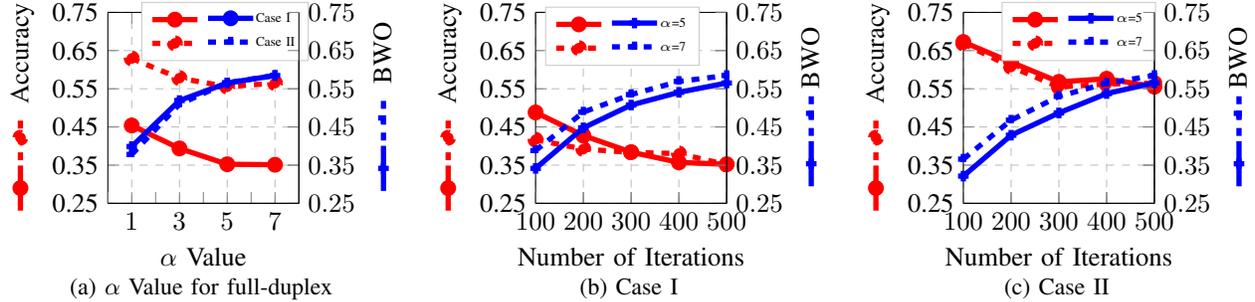

\paragraphX{\textbf{Attack Settings.}}
We test with two different settings, white-box and black-box. In all of the evaluations, the classifier is trained using adversarial training, where the attacker has full access to the defense and uses it to generate defended samples for each class in the monitored set.

\begin{itemize}
    \item \emph{White-box}: In the white-box setting, we assume that that defense uses the same network architecture for the detector as the attacker uses to perform WF attacks.
    %We define WBA where the same network architecture is used for generating adversarial traces and attacking the generated adversarial traces. In particular, the defender and attacker is assumed to have the same capability. For parameter search such as 
    We use this overly optimistic scenario only for parameter search to identify values for $\alpha$ and number of iterations, where we use DF as both the detector and the attack classifier.
    %, the white-box setting is adapted. We use DF as the primary attack classifier for most of our experiments to evaluate the attack performance in a white-box setting, as it is the state-of-the-art attack.
    
    \item \emph{Black-box}: In the black-box setting, the defender and the attacker use two different neural networks, i.e. the defender uses one model for the detector, while the attacker uses another model for performing WF attacks. We evaluate \system~in the black-box setting by using the AWF CNN model~\cite{Rimmer2018} as the detector
    and both the DF model~\cite{sirinamDF} and Var-CNN~\cite{bhat2019var} as attacker models.
    %The architecture of the detector is shown in Table~\ref{cnn_modelarch}. 
    Since DF and Var-CNN are more powerful than the simple AWF model~\cite{sirinamDF, bhat2019var}, this tests the case that the attacker has greater capabilities than are known to the defender.
    %In addition, we consider a scenario where the attacker's capability is higher than the defender. To evaluate \system~for this scenario, we generate adversarial traces using a simple CNN model and attack the generated traces with DF model for HD dataset. 
    These evaluations show the extent to which adversarial examples generated by \system~transfer to other classifiers.
    
    \item \emph{Traditional ML Attack}: We also evaluate \system~against tradition machine-learning (ML) attacks such as CUMUL~\cite{cumul}, $k$-FP~\cite{kfp}, and $k$-NN~\cite{Wang2014}, all of which reach over 90\% accuracy on undefended Tor using closed-world datasets with 100 samples per class.
\end{itemize}

\paragraphX{\textbf{Training and Hyperparameter Tuning.}}
\input{tables/hyper_para}
To perform the attacks, we use 80\% of the data for training, 10\% for validation, and the remaining 10\% for testing for each of the settings. To represent the data in the attack models, we follow the prior work~\cite{bhat2019var,sirinamDF,Wang2014,cumul,kfp} and represent the data as an 1-D vector of $+1$ and $-1$ for outgoing and incoming packets, respectively. We use a fixed length of 5000 for each instance of the class following the prior work. %For the traces defended by W-T, WTF-PAD, and \system~, we use 10,000 as the length of the instances of a class. As defenses pad dummy packets which in turn increases the total number of packets in an instance, a higher length for the defended traces will help to evaluate the attack robustness more appropriately.
%Instances that do not have 5,000 packets for undefended data and 10,000 packets for defended data are padded with zero, and the instances that have length more than 5,000 for undefended data and more than 10,000 for defended data are truncated to that particular length.
Instances that do not have 5,000 packets are padded with zero, and the instances that have length more than 5,000 are truncated to that particular length.

As black-box setting is the most realistic attack setting to evaluate a defense, we perform hyperparameter tuning on the two deep-learning based attack models: DF~\cite{sirinamDF} and Var-CNN~\cite{bhat2019var}. However, we exclude the hyperparameters choices that have been explored in prior work and did not provide any improvement in the attack accuracy. Such hyperparameters include the SGD and RMSPros optimization functions, and the tanh activation function.

We start our search with the default model and continue tuning each hyperparameter one by one. First, we perform several sets of experiments with different training epochs for both full-duplex (FD) and half-duplex (HD) with both DF~\cite{sirinamDF} and Var-CNN~\cite{bhat2019var} models. Then, the best training epoch number is used for the experiments to search for the next set of hyperparameters. Table~\ref{hyper_para} shows the hyperparameters and the choices of our hyperparameter tuning process. %The model with best hyperparameter setting for the attack using the undefended data is used to perform the attack for the defended data. 
In our tuning process, we found that the tuned DF and Var-CNN models work better for the FD traces defended by \system~and default DF and Var-CNN models work better for the HD traces defended by \system.

\paragraphX{\textbf{Top-$k$ Accuracy.}}
Most prior works have focused their analysis on Top-1 accuracy, which is normally referred to simply as the accuracy of the attack. We argue that Top-1 accuracy does not provide a full picture of the effectiveness of a defense, as an attacker may use additional insights about their target (language, locale, interests, etc.) to further deduce what website their target is likely to visit. 
%Furthermore, most WF defenses are evaluated in a simulated environment like our own. These simulators abstract away networking issues that may occur when the defense is implemented and limited by real-world network conditions. %These real-world limitations can produce imperfections when applying the defense and may expose the user to increased fingerprintability. 
 
As such, it is desirable to examine the accuracy of Top-$k$ predictions, where $k$ is the number of top-ranked classes in the prediction. In evaluations of WF, we are particularly interested in the Top-2 accuracy. A high Top-2 accuracy indicates that the classifier is able to reliably determine the identity of a trace to just two candidate websites. This is a threat to a user even when the attacker is unable to use additional information to predict the true site. The knowledge that the target {\em may} be visiting a sensitive site is actionable and can encourage the attacker to expend additional resources to further analyze their target. 

%To better evaluate the efficacy of \system~against this threat, we compute the Top-2 accuracy and compare it to that of WTF-PAD and W-T. 

%We also use DF as the primary attack classifier for most of our experiments to evaluate the attack performance in a white-box setting, as it is the state-of-the-art attack. Later, we show results using other attacks from the literature. In all of the evaluations in this section, the classifier is trained using adversarial training, where the attacker has full access to the defense and uses it to generate defended samples for each class in the monitored set. We also discuss the impact of using a less powerful detector (weaker surrogate network) in Appendix~\ref{weak_surrogate}.

%\begin{figure}[!t]
%	\centering
%	\input{figures/alpha.tex}
%    \vskip -0.3cm
%	\caption{\textit{$\alpha$ Value for full-duplex}: the accuracy and bandwidth overhead of generated samples as $\alpha$ varies. Dashed lines show the results of Case I and solid lines show the results of Case II.} % caption line
%	\label{fig:alpha_full_dup}
%\end{figure}

\paragraphX{\textbf{Target Pool.}}
%\todo[inline]{MOHSEN:  why do we consider these scenarios?}
%\todo[inline]{MI to SR:  How is that now}
\system~changes the source sample toward a target sample drawn randomly from our target pool. The detector determines whether the perturbed source sample is still in the source class. We are interested to know how the algorithm performs if we fill the target pool with instances of sites that the detector has been trained on and has not been trained on. We examine the bandwidth overhead and reduction in the attack accuracy of traces protected by \system~in these two scenarios.

\begin{itemize}
\item \emph{Case I}: We fill the target pool with instances from the Adv Set. Therefore, both source samples (${I}_{s} \in \mathcal{A} $) and target samples (${{I}^{j}}_{i} \in \mathcal{A}$ which ${T}_{i} \neq s $) are from the Adv Set. In this case, we assume that the detector has been trained on the target classes, which makes it more effective at identifying when the sample has left one class for another. This may be less effective, however, if the adversary trains on the source class but none of the other target classes used in detection.

\item \emph{Case II}: We fill the target pool with instances from unmonitored sites that are not in the Adv Set. We select the target samples (${{I}^{j}}_{i}$) from the open-world dataset. The source samples (${I}_{s}$) are from Adv Set, and we generate their defended forms. 
%The Detector Set is only used to train the {\em detector}. 
In this case, we assume the detector has not been trained on the target samples, so it may be less effective in identifying when the sample leaves the class. That may make it more robust when the attacker also trains on a different set of classes in his monitored set.
\end{itemize}

Case I may be realistic when the defender and the attacker are both concerned with the same set of sites, such as protecting against a government-level censor with known patterns of blocking, e.g. based on sites it blocks for connections not protected by Tor. Case II is a more conservative estimate of the security of \system~and thus more appropriate for our overall assessment.

We generate defended samples with various settings. We vary $\alpha$ to evaluate its effect on the strength of the defended traces and the overhead. 
%We measure the detectability of the defended samples by applying the DF attack~\cite{sirinamDF} on them. 
%The WF defenses protect the traffic traces by adding dummy packets to the traffic, this increases the number of packets in the traces. Therefore, the number of packets in the defended traces is larger than in their undefended forms. We thus increase the input size of the DF attack and .
We also vary the number of iterations required to generate the adversarial traces. Each iteration moves the sample closer to the target, improving the likelihood it is misclassified, but also adds bandwidth overhead. 
%The bandwidth overhead gets increased with the increase of iteration. On the other hand, the accuracy gets better and better. However, we cannot make the changes to an unreasonable number of iterations. We want to control the bandwidth overhead to a limit with an acceptable attack accuracy. 

%%% prob threshold
%\input{figures/prob_threshold_fd}
%\input{figures/prob_threshold_hd}

\begin{figure*}[!t]
\centering
    \begin{subfigure}[h]{0.5\textwidth} \centering
        \includegraphics[scale=0.4]{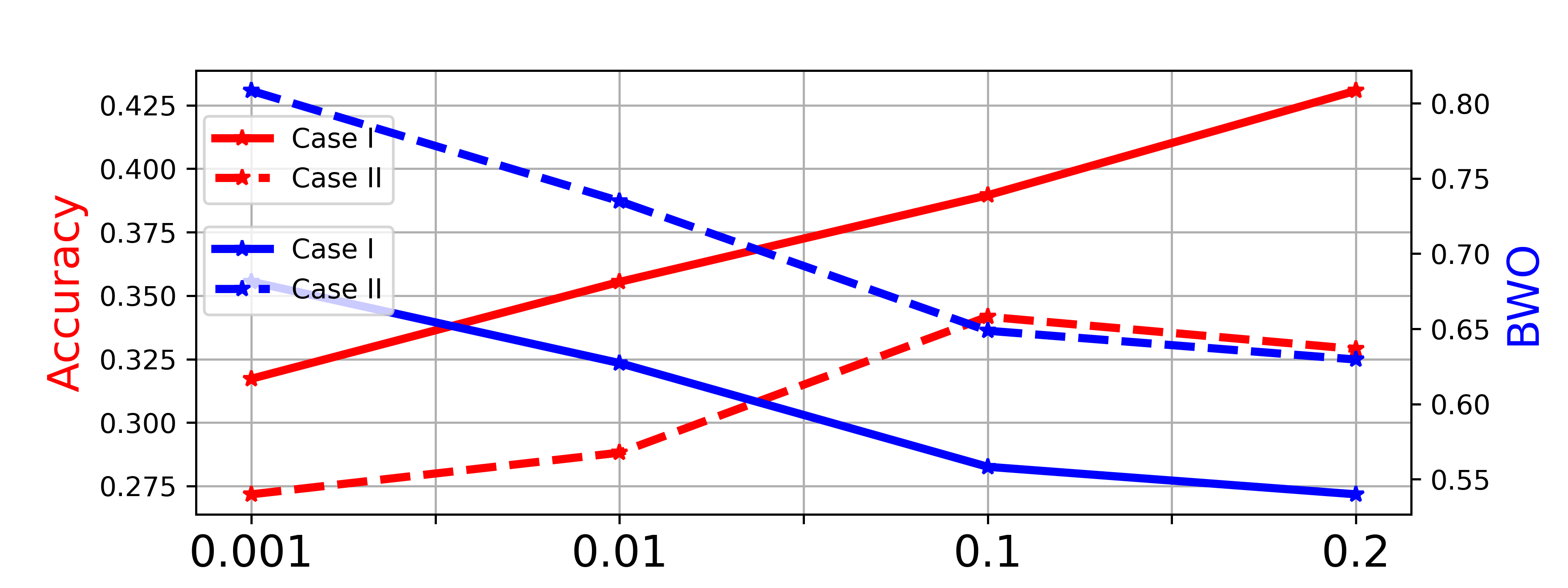}
        \vskip -0.2cm
	\caption{Confidence Threshold Value ($\tau_c$)} % caption line
	\label{fig:prob_threshold_hd}
    \end{subfigure}%
    ~
    \begin{subfigure}[h]{0.5\textwidth} \centering
        \includegraphics[scale=0.4]{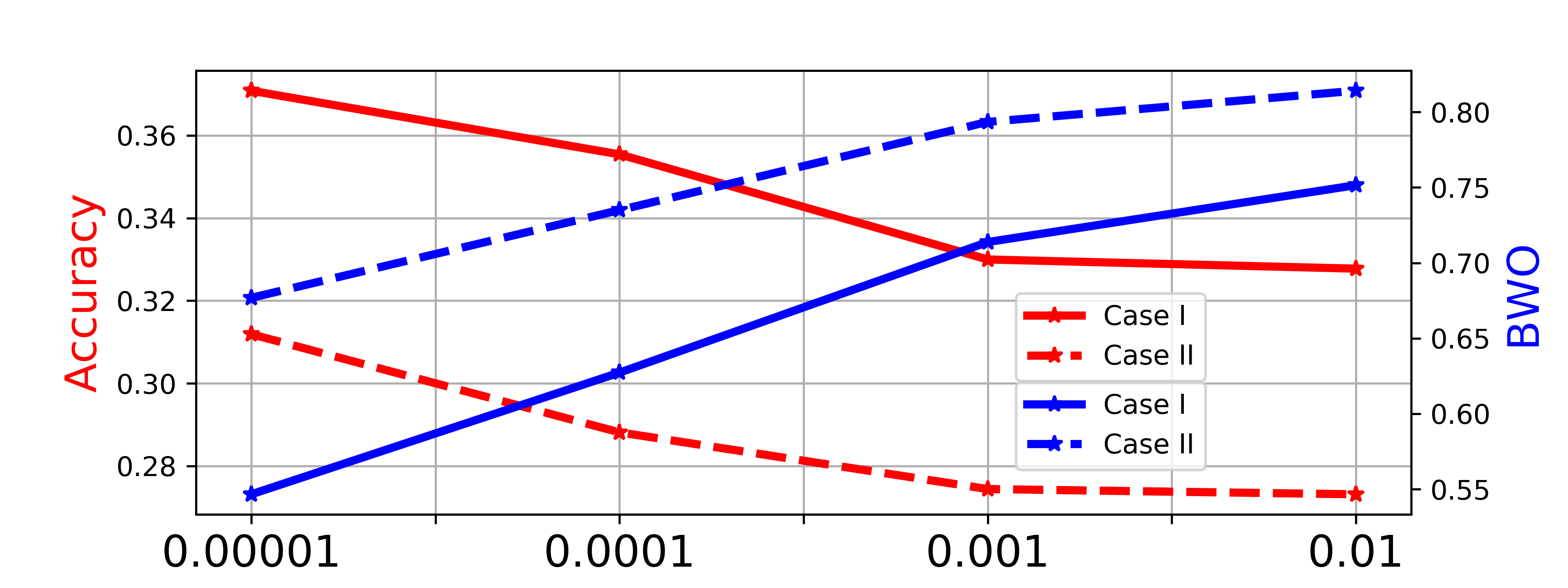}
        \vskip -0.2cm
        \caption{Perturbation Threshold Value ($\tau_D$)}
        \label{fig:change_threshold_hd}
    \end{subfigure}%
\vskip -0.2cm    
\caption{\textit{Half-duplex}: the attacker's accuracy and the bandwidth overhead (BWO) of the generated samples as we vary the probability threshold value (Figure~\ref{fig:prob_threshold_hd}) and perturbation threshold value (Figure~\ref{fig:change_threshold_hd}).}
% Dashed lines show the results of Case I and solid lines show the results of Case II.
\label{fig:threshold_hd}\end{figure*}

\subsection{Tuning on Full-Duplex Data}
\label{tuneFD}
The full-duplex version of \system~is the easier to deploy and leads to lower latency costs than the half-duplex version~\cite{Wang2017}, so we lead with the full-duplex results. We use the white-box setting for simplicity.

\paragraphX{\textbf{Choice of Threshold Values.}}
There are two thresholds values in \system~algorithm: i) a confidence threshold value ($\tau_c$) that limits the confidence on the generated trace belonging to the correct class, and ii) a perturbation threshold value ($\tau_D$) that limits the amount of change in a generated trace. Intuitively, the smaller the value of $\tau_c$, the less likely it is that the attacker could succeed, at the cost of higher average bandwidth overhead. By contrast, changing $\tau_D$ directly manipulates the maximum bandwidth overhead, where lower $\tau_D$ would reduce bandwidth cost but also typically result in an improved chance of attack success. 
%Hence, these is a trade-off between bandwidth overhead and attack accuracy in choosing threshold values ($\tau_c$ and $\tau_D$).

While we experiment with different $\tau_c$ values, we fix the $\tau_D = 0.0001$, $\alpha = 5$, and number of iterations to 500. We can see from Figure~\ref{fig:prob_threshold_fd} that lower values of $\tau_c$ lead to lower accuracy and higher bandwidth. $\tau_c = 0.01$ provides a good tradeoff point, so we select it for our following experiments.
%reduces the attack accuracy with an increase of bandwidth overhead. We can see that $\tau_c = 0.01$ provides a better trade-off between attack accuracy and bandwidth overhead. Hence, we select  $\tau_c = 0.01$ for our next set of experiments.

Following the same procedure, we perform experiments with different $\tau_D$ values. As expected, and as shown in Figure~\ref{fig:change_threshold_fd}, lower $\tau_D$ values lead to lower bandwidth but higher attack accuracy. We choose $\tau_D = 0.0001$ as a good tradeoff for our following experiments.

%\paragraphX{Input Length Selection}
%\input{tables/input_length.tex}
\paragraphX{\textbf{Choice of $\alpha$.}}
Figure~\ref{fig:alpha_full_dup} shows the bandwidth overhead and attack accuracy of full-duplex data with respect to $\alpha$ values for both Case I (solid lines) and Case II (dashed lines) with 500 iterations. %We explain later about the process of selecting the number of iterations to generate adversarial traces. In our experiments, we notice that odd number for $\alpha$ value works better and we continue with different odd number starting from 1.  
As expected, the bandwidth overhead increases and the attack accuracy decreases as we increase $\alpha$, with longer steps towards the selected targets.
%Larger $\alpha$ values create longer steps toward the target samples lead to higher bandwidth overhead.
For Case I, the adversary's accuracy against \system~with $\alpha$=5 and $\alpha$=7 are both 35\%, but the bandwidth overhead is lower for $\alpha=5$ at 56\% compared to 59\% for $\alpha=7$. For Case II, the adversary's accuracy and the bandwidth overhead are both slightly lower for $\alpha$=5 than that of $\alpha$=7. From these findings, we fix $\alpha$=5 for our experiments.
%the rest of our other experiments in both cases.

We also observe that Case I leads to lower accuracy and comparable bandwidth overhead to Case II. When $\alpha$=5 and $\alpha$=7, the attack accuracies for Case I are at least 20\% lower than that of Case II. Therefore, as expected, picking target samples from classes that the detector has been trained on drops the attacker's accuracy. 

\paragraphX{\textbf{Number of Iterations.}}
Figure~\ref{fig:adv-DF-acc_full_dup} shows the trade-off between the accuracy and bandwidth overhead with respect to the number of iterations to generate the adversarial traces. As mentioned earlier, increasing the number of iterations also increases the number of packets (overhead) in the defended traces. %It is obvious that if we add a lot of bandwidth, we can defeat the attack. But that would not be a realistic defense. 
We vary the number of iterations from 100 to 500 for both Case I (Figure~\ref{fig:adv-cw-DF-acc}) and Case II (Figure~\ref{fig:adv-open-DF-acc}) to see their impact on the overhead and the accuracy rate of the DF attack.

For Case I, we can see that the DF attack accuracy for both 400 and 500 iterations is 35\% when $\alpha$=5, while the bandwidth overheads are 54\% and 56\%, respectively. For $\alpha=7$, the attacker's accuracy is higher and the bandwidth costs are higher.
%The attack accuracy is 3\% higher when $\alpha$=7 for 400 iterations. 
%For $\alpha$=7, it is 2\% higher for both 400 and 500 iterations. This is another evidence that $\alpha$=5 is better to generate adversarial traces.
%
For Case II, using $\alpha$=5 leads to 57\% accuracy with 53\% bandwidth overhead for 400 iterations and 55\% accuracy and 56\% bandwidth overhead for 500 iterations. From these findings, we fix the number of iterations to 500 for our experiments.
%As a defender we want to reduce the attack accuracy as much as possible with an acceptable amount of bandwidth overhead. From the figures, we can interpret that 500 iterations provides reduced accuracy than 400 iterations with comparable amount of bandwidth overhead. Hence, we keep the number of iterations to 500.

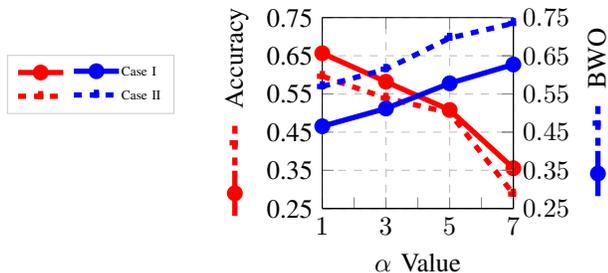
\begin{figure}[t]
	\centering
	\input{figures/C1_half_acc_bw.tex}
    \vskip -0.2cm
	\caption{\textit{Half-duplex}: attacker accuracy and bandwidth overhead (BWO) as $\alpha$ varies.} %Dashed lines show Case I and solid lines show Case II.} % caption line
	\vskip -0.3cm
	\label{fig:alpha_half_dup}
\end{figure}

\begin{figure}[!t]
    
	\centering
	\input{figures/C1_half_its.tex}
    \vskip -0.2cm
	\caption{\textit{Half-duplex}: attacker accuracy and bandwidth overhead (BWO) as the number of iterations varies.} % caption line
	\vskip -0.3cm
	\label{fig:alpha_half_its}
\end{figure}
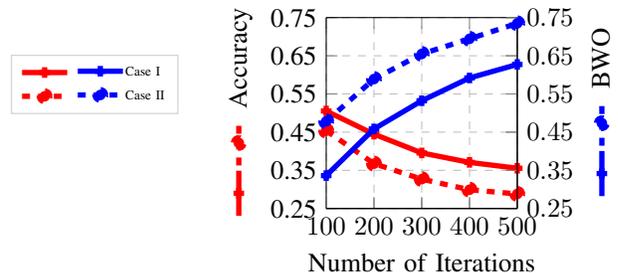

\subsection{Tuning on Half-Duplex Data} \label{exp_results_half_dup}

Using half-duplex communication increases the complexity of deployment and adds latency overhead~\cite{Wang2017}, but it offers more precise control over burst patterns such that we can achieve reduced attacker accuracy as shown in the following white-box results.

\paragraphX{\textbf{Choice of Threshold Values.}}
We can see from Figure~\ref{fig:prob_threshold_hd} and \ref{fig:change_threshold_hd} that $\tau_c = 0.01$ and $\tau_D = 0.0001$ provide better trade-off between attack accuracy and bandwidth overhead for HD data as well. The attack accuracies are 35\% and 29\%, and bandwidth overheads are 73\% and 63\% for case I and case II, respectively. Hence, we select $\tau_c = 0.01$ and $\tau_D = 0.0001$ for our next set of experiments.

\paragraphX{\textbf{Choice of $\alpha$.}}
As seen in Figure~\ref{fig:alpha_half_dup} (all for 500 iterations), the lowest accuracy rates are 35.5\% and 28.8\% for Case I and Case II, respectively, when $\alpha$=7. The bandwidth overheads are 62.7\% and 73.5\% for Case I and Case II, respectively. When $\alpha$=5, the attack accuracy is 50\% for both Case I and Case II with bandwidth overheads of 57\% and 69\%, respectively. As expected, higher $\alpha$ values mean lower attacker accuracies at the cost of higher bandwidth. Additionally, as in the full-duplex setting, both bandwidth overhead and accuracies are lower in Case I than Case II. From these findings, we set $\alpha=7$ for our experiments.
%As a defense, the reduced accuracy is more advisable even with a little more cost of bandwidth. Though $\alpha$=5 provides moderate reduced bandwidth overhead, the attack accuracy are higher for both cases compared to $\alpha$=7. Hence, for the half-duplex data, $\alpha$=7 is the preferable choice. All of the results shown in Figure~\ref{fig:alpha_half_dup} are based on 500 iterations.

%We can see that Case I provides lower bandwidth overhead and lower detectability than Case II. This means that picking target samples from classes that the detector trained on reduces both attack accuracy and bandwidth overhead. For Case II, $\alpha$=5 provides us with 57\% accuracy with 53\% bandwidth overhead for 400 iterations. For 500 iterations, it provides 55.5\% accuracy and 56.5\% bandwidth overhead. 
%As a defender we want to reduce the attack accuracy as much as possible with an acceptable amount of bandwidth overhead. We can understand from the graphs and the explanations that iterations=500 provides reduced accuracy than iterations=400 with comparable amount of bandwidth overhead. Hence, we keep the number of iterations to 500.

\paragraphX{\textbf{Number of Iterations.}}
Figure~\ref{fig:alpha_half_its} shows the trade-off between attacker accuracy and bandwidth overhead with the number of iterations for $\alpha$=7. We vary the number of iterations from 100 to 500 for both Case I and Case II. For Case I, the accuracy is 35.5\% with 62.7\% bandwidth overhead with 500 iterations. With 400 iterations, the accuracy is 37\% with 59\% bandwidth overhead. For Case II, with 500 iterations, we can get the lowest attack accuracy of 28.8\%, with a cost of 73.5\% bandwidth overhead. From these findings, we set the number of iterations to 500 for our experiments.
%We can observe that the bandwidth overhead for Case I is always lower than Case II. 

\if 0
\begin{figure}[t!]
\centering
  \includegraphics[scale=0.50]{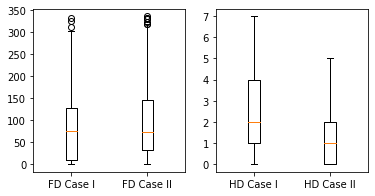}
  \caption{\textbf{Site-5:} Number of Target Changes in both HD and FD.}
  \vskip -0.3cm
  \label{fig:site_5_total_target_change}
\end{figure}
\fi

%\paragraphX{\textbf{Number of Target Changes.}} 
%In Figure~\ref{fig:site_5_total_target_change}, we show the number of times the target class changes during the search for an adversarial example for all the instances of a particular website (labeled as site 5 in both the HD and FD datasets). We see that the full-duplex (FD) traces require many more target changes than the half-duplex (HD) traces. This illustrates a significant difference in how the \system~algorithm works between FD and HD, though the overall results for both systems are similar. 

%\input{tables/top_2.tex}
%\input{tables/simple_CNN.tex}

%%%%%%%%%%%%%%%%%%%%%
\subsection{Results Analysis}
\label{results_analysis}

%Robust randomness to pick a target sample makes \system~resistant to adversarial training.

%We show a comparison of attack accuracies of \system~against DF~\cite{sirinamDF}, AWF~\cite{Rimmer2018}, CUMUL~\cite{cumul}, $k$-FP~\cite{kfp}, and $k$-NN~\cite{Wang2014} attacks  for both Case I and Case II in Table~\ref{tab_results_analysis}. We also show the comparison of the performance of \system~with two state-of-the-art defenses WTF-PAD~\cite{wtf-pad} and Walkie-Talkie (W-T)~\cite{Wang2017}. In Table~\ref{tab_top2_acc}, we show the comparison of Top-2 accuracy of \system~, WTF-PAD defense, and W-T defense.

We lead our analysis with bandwidth overhead followed by the analysis of white-box setting, black-box setting, and traditional ML attacks. We extend our analysis with a discussion of Top-$k$ accuracy. The analysis is based on the best parameters ($\tau_c$, $\tau_D$, $\alpha$ and number of iterations) found in Section~\ref{tuneFD} and \ref{exp_results_half_dup} where $\tau_c = 0.01$, $\tau_D = 0.0001$, and $\alpha = 5$ and $\alpha = 7$ for FD and HD datasets, respectively. The number of iterations are 500 for both datasets. In addition, we include the investigation of multiple-round attacks on \system. To compare \system~with other defenses, we selected two state-of-the-art lightweight defenses: WTF-PAD~\cite{wtf-pad} and W-T~\cite{Wang2017}. To generate defended traffic, we simulated WTF-PAD on our FD dataset, as it works with full-duplex traffic, while we simulated W-T on our HD datasets, since W-T uses half-duplex traffic. Thus, in our experiments, the WTF-PAD dataset has 95 sites with 518 instances each and W-T has 82 sites with 360 instances -- note that we use just the {\em Adv Set} $\mathcal{A}$ from the HD dataset for a fair comparison with \system.

For our W-T simulation, we used half-duplex data from Sirinam et al.~\cite{sirinamDF}.\footnote{Tao Wang's website (\url{http://home.cse.ust.hk/~taow/wf/}, accessed Sep. 10, 2020) mentions that Sirinam et al.'s W-T browser, which is designed to collect half-duplex data, is a better implementation.} There are modest differences between this data and the data used by Wang and Goldberg~\cite{Wang2017}. Perhaps most importantly, Sirinam et al.’s implementation uses a newer version of Tor. Also, Sirinam et al.’s dataset is from websites in 2016 rather than 2013, which we have observed makes a significant difference in their distributions. These differences likely account for the differences in reported bandwidth in our findings and those of Wang and Goldberg. 
%The molding process in our W-T simulation is based on perfect symmetric collision. We divide 82 sites into two parts: sensitive (41 sites) and non-sensitive (41 sites). Perfect symmetric collision ensures that site 0 is molded with site 41 and vice versa for each of the sensitive sites. The W-T bandwidth in our dataset is higher than that of the W-T paper. We hypothesize that data from the half-duplex browser using a newer version of Tor, and a different distribution of data are the potential reasons for that.

\paragraphX{\textbf{Bandwidth Overhead.}} 
%\subsubsection{Bandwidth Overhead}
%Bandwidth overhead is the amount of additional packets that are added to the undefended traces to make the traffic of each site indistinguishable. However, it comes at the cost of extra communication overhead that can congest the network traffic in Tor. Thus, it is preferable to keep the bandwidth overhead as low as possible since increased bandwidth overheard will hurt the user's experience. 
As described in Section~\ref{genadvtraces}, we designed \system~to minimize packet addition, and thus bandwidth overhead, by keeping the amount of change in a trace under a threshold value of 0.0001. %Keeping the amount of change at a minimum keeps the bandwidth overhead as low as possible. 
%We show comparison against WTF-PAD and W-T because these are the two lightweight and state-of-the-art defenses, while the other proposed WF defenses have two-to-three times higher delays and bandwidth overheads. 
From Table~\ref{tab:black-box}, we can see that for full-duplex (FD) network traffic, the bandwidth overhead in Case I and Case II of \system~are the same at 58\%, which is 6\% and 14\% lower than WTF-PAD and W-T, respectively. For half-duplex (HD) \system, the bandwidth overhead is 62\% for Case I and 70\% for Case II. 
%In all cases, then, the bandwidth costs are in line with or modestly better than WTF-PAD and W-T and acceptable for use in Tor.

%In half-duplex (HD) network traffic, Case II requires at least 10\% higher bandwidth overhead than Case I. In terms of bandwidth overhead, Case I is better for both FD and HD network traffic, although not to a significant degree. Overall, Case II for HD network traffic provides the strongest defense (28.8\% accuracy) but requires the highest cost (73.5\% bandwidth overhead) for any of the lightweight defenses. 
%us the lowest accuracy, 28.8\%, among all the cases of our defense, though costs  73.5\% additional bandwidth, which is more than any other cases

%In summary, our bandwidth overheads are better than WTF-PAD and W-T in most of the cases, except one. This amount of bandwidth overhead can be acceptable in comparison to other WF defenses for Tor. 

\input{tables/results_analysis.tex}
\paragraphX{\textbf{White-box Setting.}}
In the white-box setting, shown in Table~\ref{tab_results_analysis}, \system~is highly effective at fooling the attacker. In Case I, the attacker gets less than 40\% accuracy for both FD and HD traffic. For Case II, the attacker actually has a lower accuracy of 29\% in the HD setting, but reaches 55\% in the FD setting.

\input{tables/surrogate}

\paragraphX{\textbf{Black-box Top-1 Accuracy.}} 
%In the black-box setting, the defender's detector and the attacker's classifier are two different neural networks, where the detector model (AWF) is weaker than the attacker's model (DF or Var-CNN). We treat this case as the most likely one and more appropriate for comparison with other defenses. 
In the more realistic black-box setting, our results as shown in Table~\ref{tab:black-box} indicate that \system~is an effective defense. For Case I and Case II, 
%the respective bandwidth overheads are 62\% and 73\%, while 
the respective attack accuracies are at most 42\% and 62\%.
%(Var-CNN on Case II, full duplex).
In comparison, Var-CNN achieves 90\% closed-world accuracy against WTF-PAD and 44\% against W-T. So the effectiveness of \system~in terms of Top-1 accuracy falls between these two defenses. 

\paragraphX{\textbf{Traditional ML Attacks.}} As shown in Table~\ref{weak_detector}, the highest performance for any of CUMUL, $k$-FP, and $k$-NN against \system~was 32\% (Case II, Full Duplex). \system~is competitive with or better than both WTF-PAD and W-T for all three attacks. This shows that the effectiveness of \system~is not limited to protecting against DL models, but also against traditional ML.
%The CUMUL attack can achieve at best 32\% accuracy against FD network traffic of Case I. In all the other cases, the CUMUL attack accuracies against \system~are closer to 22\% or lower. CUMUL attack accuracies against \system~are significantly lower than against WTF-PAD and close to the accuracy against W-T (see Table~\ref{weak_detector}). 
%It is important to note that the CUMUL attack is also unsuccessful against our defense.
%For the $k$-FP attack, the results are similar, with W-T slightly outperforming one case of \system. For $k$-NN, \system~outperforms the other two defenses.
%Though \system~is more effective than WTF-PAD against $k$-FP, for one case it is slightly higher than W-T. However, $k$-NN performs worst against our defense in all the cases than WTF-PAD and W-T.
%These results show that the effectiveness of \system~is not limited to protecting against DL white-box and black-box attacks, but also against traditional ML attacks, which get at most 32\% accuracy in all cases.

%It is also notable that both Case I and Case II provide consistencies between the attack accuracies for both FD and HD network traffic against all the attacks. 
%It is interesting that all the attacks accuracies against Case I FD is lower than that of Case I HD. On the other hand, attack accuracies against Case II FD are higher than for Case II HD. Overall, the results demonstrate that \system~is effective and has lower attack performance compared to the state-of-the-art WF defenses, WTF-PAD and W-T. 

\paragraphX{\textbf{Top-k Accuracy.}}
%\subsubsection*{Top-N Accuracy}
Our results in Table~\ref{tab:black-box} show that \system~is somewhat resistant to Top-2 identification, with an accuracy of 72\% in the worst case. On the other hand, \mbox{W-T} struggles in this scenario with 97\% Top-2 accuracy by DF, as its defense design only seeks to provide confusion between two classes. In addition, Var-CNN attains 95\% Top-2 accuracy against WTF-PAD. 
%\todo{round all the results in the table correctly. 89.7 is 90, not 89}
%90\% Top-2 accuracy. 
This Top-2 accuracy of \system~indicates a notably lower risk of de-anonymization for Tor users than WTF-PAD or W-T.

%We also analyze the Top-3 to Top-10 accuracy and discuss why \system~is good at restricting the Top-$k$ accuracy in Supplementary Material.% Appendix~\ref{top_k_analysis}.
%\input{appen_C.tex}

\begin{figure}[t!]
\centering
  \input{figures/top_k_all}
  \caption{\textbf{Black-Box Top-$k$ Accuracy.} DF accuracy for different values of $k$ against \system.} 
  %The $x$-axis represents the Top-$k$ and $y$-axis represents the attack accuracy
  %Red and Green solid line represents the results for full-duplex dataset and dashed line represents the results for half-duplex dataset.}
  \label{fig:top_k_10}
\end{figure}
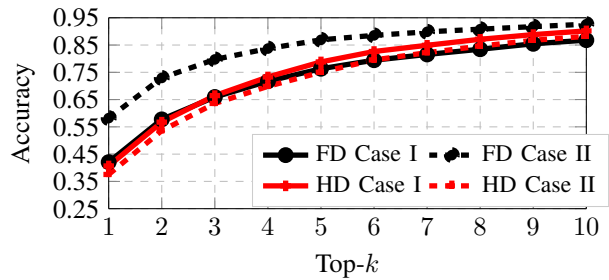

%\vspace{-0.5cm}
%\begin{figure}[ht!]
%\centering
%  \includegraphics[scale=0.55]{figures/HDFD_BOX_Total.png}
%  \caption{Total Number Target Changes for both HD and FD dataset.}
%  \label{fig:total_target_change}
%\end{figure}

In addition to Top-2, we analyzed the Top-10 accuracy of DF against \system. Figure~\ref{fig:top_k_10} shows that \system~can limit the Top-10 accuracies of full-duplex (FD) data to 87\% and 92\% for Case I and Case II, respectively. For half-duplex (HD), Top-10 accuracies are 90\% and 88\% for Case I and Case II. Overall, the worst-case Top-10 accuracy is about the same as the Top-2 accuracy against WTF-PAD, while the worst-case Top-10 accuracy is significantly better than the Top-2 accuracy for W-T.
\subsection{Intersection Attacks}
\label{multiroundattacks}

\input{tables/multi_round}
In this section, we evaluate the effectiveness of \system~in a  scenario in which the adversary assumes that the user is going to the same site regularly (e.g. every day), and that the adversary is in a position to monitor the user's connection over multiple visits. Both of these assumptions are stronger than in a typical WF attack, but they are common in the literature on attacking anonymity systems, such as \emph{predecessor attacks}~\cite{wright2004predecessor,wright2008passive} and \emph{statistical disclosure attacks}~\cite{danezis2003statistical,mallesh2011analysis}. In this setting, the attacker can leverage weaker information such as top-10 classification results to eventually deanonymize a user who uses Tor to visit the same site every day. 

While more sophisticated statistical methods could be adopted~\cite{danezis2003statistical,mallesh2011analysis}, we have limited data for this purpose and instead use a simple \emph{intersection attack} to shed light on the information available to the attacker. In an intersection attack, the adversary examines the top-$k$ results of a WF attack applied to each of the user's Tor connections. Treating each result as a set of $k$ sites, he takes the intersection of the sets to eliminate sites that do not appear every day. If the results consistently identify a single site as being in the top-$k$ results, then one may speculate that this site is in fact where the user is going online using Tor every day. 

If the WF defense is very consistent, in that the same $k$ sites are in the top-$k$ results every day, then this kind of intersection attack will not reduce the security of the defense over time. In that sense, W-T provides an assurance that the top-2 accuracy holds for this stronger attack model.

For \system, however, there is no direct assurance. We thus test the intersection attack over a period of five rounds (i.e. five days). We model multiple rounds by randomly selecting a set of test instances with the same label from our dataset.
%As the top-10 accuracy of \system~is around 90\%, an attacker can perform multiple rounds of attacks and an intersection of the predictions after multiple rounds may eventually reveal the correct class. Based on this concern, we design our experiment to perform a five-round attack. 
For each round, we compute the top-10 predicted labels ($T_{10}$) and take the intersection of that set with the
intersected labels ($L_{int}$) computed in the previous round. 
%$L_{int}$ can be considered as a prior for the subsequent rounds. %For the first round, however, the intersected labels ($L_{int}$) and the top-10 labels ($T_{10}$) are the same as there is no prior for the first round. 
Mathematically, this attack process can be expressed as $\nabla L_{int}^{n} = T_{10}^{n} \cap L_{int}^{n-1}$ for each attack round $n>1$.

The metrics we use for evaluating \system~in this scenario are: i) {\em absolute success}, ii) {\em absolute failure}, and iii) {\em mean intersection}.
After five rounds, if $L_{int}$ consists of only the correct class, we call this an absolute success. Absolute failure is the case where $L_{int}$ is empty, or the correct class is not in $L_{int}$. Finally, for cases where $L_{int}$ contains two or more classes including the correct class, we take the average size of $L_{int}$ as the mean intersection.

We use the tuned DF model for \system~FD data and the default DF model for \system~HD data for these experiments. The results are shown in Table~\ref{multiround}. Absolute successes for FD are 27\% and 36\% for case I and case II, respectively; they are below 20\% for both HD cases. The absolute failure rates are slightly above 50\% for FD case I and both HD cases, though just 20\% for FD case II. Mean intersections for FD are 2.86 and 2.60 for case I and case II, respectively, and 4.12 for both HD cases. 

From this, we see that the intersection attack is only moderately helpful for the attacker against \system. Even when the attacker can assume that the user visits the same site regularly, the top-10 labels do not include the real site in a significant fraction of cases. Further, there are often multiple sites in the intersection set. Thus, while \system~does not provide assurance against attacks over multiple rounds of observation by a determined attacker, it is also not highly vulnerable to such an attack. 

%From this set of results, we can infer that although multiple-round can provide an edge to the attacker, the error rate is still above 50\% for the most of the settings. However, there can be different ways to take advantage of this multiple-round attacks, we leave this investigation as part of future work.

%%%%%%%%%%%%%%%%%%%%%
\subsection{Information Leakage Analysis}
\label{infoleak_analysis}
%\matt{Figure: please change caps to "WTF-PAD"}
%\matt{What variant of \system are we evaluating here? please specify.}

\begin{figure}[!t]
    \centering
  \includegraphics[scale=0.55]{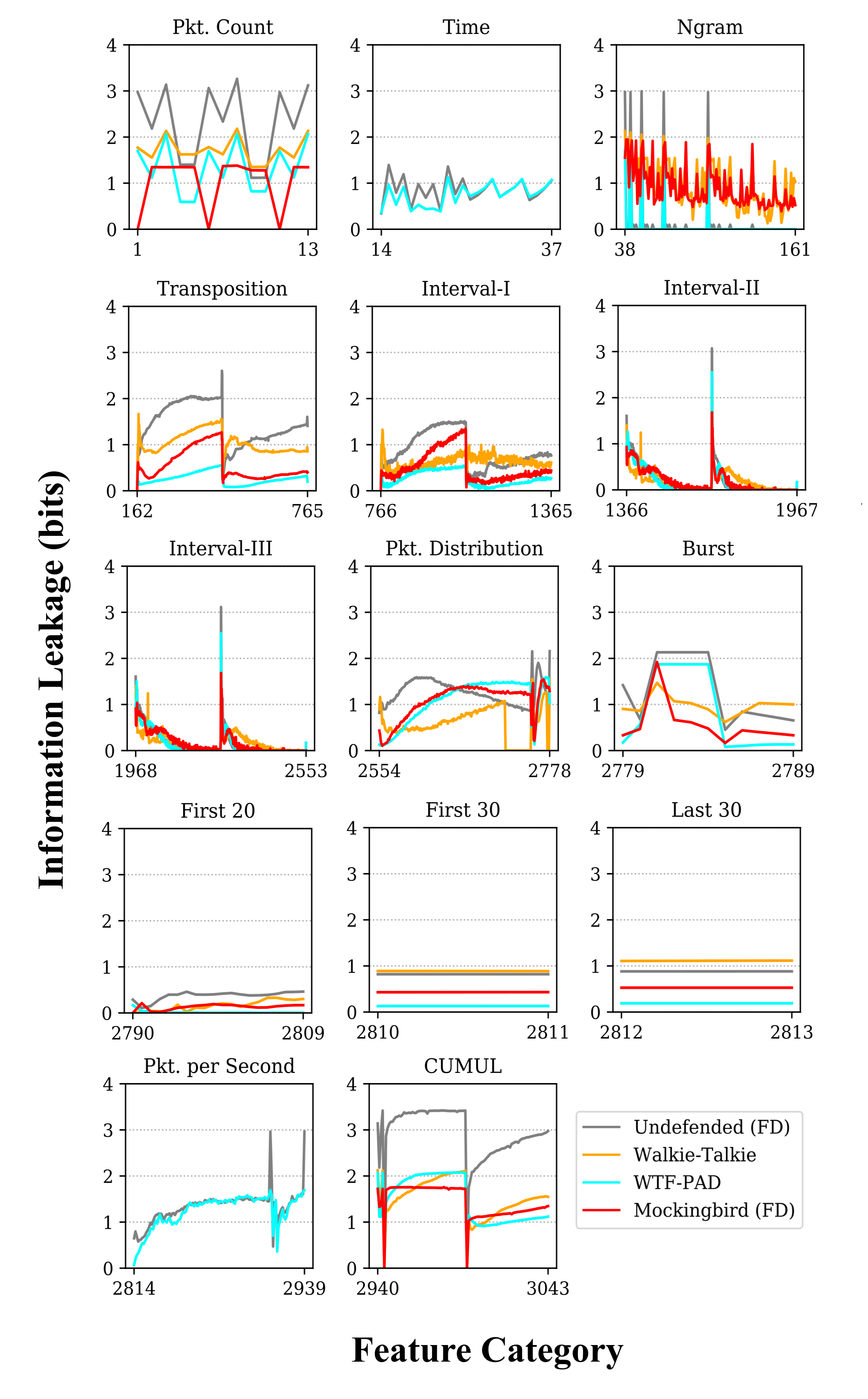}
  \caption{Individual feature information leakage.}
  \label{fig:infoleak}
\end{figure}

Recent works have argued that classification accuracy is not a complete metric to evaluate the objective efficacy of a WF defense~\cite{infoleak,cherubinbayes}. We thus adopted the WeFDE information leakage estimation technique~\cite{infoleak} to evaluate \system. WeFDE allows us to determine how many bits of information each feature leaks for a given defense. The authors of the WeFDE paper made their code available to us; for speed and memory requirements, we re-implemented most of the code ourselves and validated our common results against theirs. We perform information leakage analysis on undefended full-duplex traffic and defended traffic for 3043 manually defined features spread across 14 categories. The defenses we examine are WTF-PAD, W-T, and the full-duplex Case I variant of \system. The leakage for each feature and defense is shown in Figure~\ref{fig:infoleak}. Our {\em Timing} and {\em Pkt. per Second} categories do not include W-T or \system~measurements, as the simulations for these defenses are unable to produce accurate timestamp estimations.

In general, we find that the information leakage of \system~to be comparable to the other defenses. We find that any given feature leaks at most 1.9 bits of information in \system. W-T has similar maximum leakage, with at most 2.2 bits per feature, while WTF-PAD leaks at most 2.6 bits. The maximum amount of leakage seen for the undefended traffic was 3.4 bits, nearly twice that of \system. \system~seems to be most vulnerable to the $N$-gram class of features, which is not surprising, as these features seem to be effective for all traffic defenses we examined. On the other hand, it is quite effective against packet count features, which was previously the most significant class of features for the undefended traffic and also useful against W-T. Additionally, \system~shows notable improvements over W-T in the {\em Transposition}, {\em First 30}, {\em Last 30}, and {\em Burst} features, while W-T is better than \system~in the {\em Pkt. Distribution} feature. 

Overall, the results of our information leakage analysis are largely consistent with what we see in our accuracy measurements.
%: W-T and \system~ achieve similar performance, with \system~ slightly edging out W-T.

%% file: 5_1_Datasets.tex
\subsection{Datasets}\label{dataset}

We apply \system~to generate adversarial examples on the traffic traces at the burst level. %We define the bursts as the sequence of the consecutive packets in the same direction.
We can get the burst sequence of the traffic traces from both full-duplex (FD) and half-duplex (HD) communication modes. %The traffic traces in~\cite{Cherubin2017, kfp, sirinamDF,Wang2014} were collected as the communications between clients and servers are in the full-duplex mode. %This means that the incoming (outgoing) bursts might be interrupted by outgoing (incoming) bursts because before receiving the full response from the previous request, the client may send new requests.
%We can use burst sequence of this full duplex traffic and generate adversarial traces for them. In addition, we can use the burst sequence of half-duplex traffic traces to generate the adversarial traces, the burst sequence of half-duplex communication is more representative of the burst in the traffic.. %We need to collect the traffic on the half-duplex communication.
Walkie-Talkie (W-T)~\cite{Wang2017} works on HD communication, and it finds the supersequences on the burst level. In our experiments, we use burst sequences for both FD and HD datasets.

\paragraphX{Data Source:} We use both the closed-world (CW) and open-world (OW) FD and HD traffic traces provided by Sirinam et al.~\cite{sirinamDF}. The websites classes from the monitored set are from the top sites in Alexa~\cite{alexa}.

\paragraphX{Preprocessing Data:} In our preprocessing phase, we filter the data by removing any instances with fewer than 50 packets and the instances that start with an incoming packet, since the client should send the first packet to start the connection.

\paragraphX{Full-Duplex (FD):} %We use the FD traffic traces provided by Sirinam et al.~\cite{sirinamDF}% collected a reasonably large dataset of the FD traffic traces.
%Their dataset has 95 classes with 1000 instances each. 95 sites are from the top 100 sites in Alexa~\cite{alexa}.  After this processing, we end up with 518 instances for each site. We also use the FD open-world (OW) dataset from Sirinam et al.~\cite{sirinamDF} in our experiments which has 40,716 different sites with 1 instance each. We process the OW dataset as well.
The CW dataset contains 95 classes with 1000 instances each. After preprocessing, we end up with 518 instances for each site. The OW dataset contains 40,716 different sites with 1 instance each.

\paragraphX{Half-Duplex (HD):} %We use the HD dataset provided by Sirinam et al.~\cite{sirinamDF}. This dataset contains 100 sites which are also from top 100 sites in Alexa.com~\cite{alexa}, with 900 instances for each class. We preprocess this dataset in the same way we processed FD dataset. %We cleaned their data and removed the traces shorter than 50 packets as well as traces which began with incoming packets.
%After preprocessing, we ended up with 83 classes with 720 instances per class. Moreover, they also have OW data of 40,000 sites with 1 instance each. We also process OW dataset. We use both of these HD closed-world (CW) and OW datasets for our HD evaluations.
The CW dataset contains 100 sites with 900 instances each. After preprocessing, we ended up with 83 classes with 720 instances per class. The OW data contains 40,000 sites with 1 instance each.

\vskip 0.2cm
%\todo[inline]{MI: if we don't do any open-world evaluation we should remove them from here}
%\todo[inline]{Saidur to MI: But we need open world data for Case II, right?}
%Further preprocessing is required before we can use these datasets to evaluate our defense. The burst sequences vary in size between different visits and sites. 
An additional consideration is that we must use a fixed size input to our model~\cite{sirinamDF}. 
%Thus, we must determine an ideal input size and adjust our dataset to this size. To do this, 
To find the appropriate size, we consider the distribution of burst sequence lengths within our datasets. Figure~\ref{fig:cdf-burst} shows the CDF of the burst sequence lengths for both the HD and FD datasets. More than 80\% of traces have fewer than 750 bursts for the HD CW dataset, and more than 80\% of the traces for the CW FD dataset have fewer than 500 bursts. % We next applied our adversarial defense to the half-duplex and full-duplex datasets. 
We found that using 1500 bursts on both HD and FD datasets provides just 1\% improvement on accuracy for the DF attack compared to using 750 bursts.%a simple CNN model (with three convolutional layers followed by two fully-connected layers). %In addition, the accuracy rates of the full-duplex dataset with 750 bursts and 1500 bursts are 91\% and 93\%.
%The difference between the performance with the full-size burst length and its short form (750 bursts) would thus appear to be negligible. T
To decrease the computational cost for generating adversarial examples, we use an input size of 750 bursts for both the FD and HD datasets. Note that the attacker in our evaluations uses 10,000 packets rather than bursts.

\begin{figure}[t!]
	\centering
	\input{figures/both_cdf_bursts.tex}
	\caption{CDF of the number of bursts in the full-duplex and half-duplex traces.} % caption line
	\label{fig:cdf-burst}
\end{figure}

In our evaluation, we need to address the needs of both \system~and the adversary to have training data. We thus break each dataset (full-duplex (FD) and half-duplex (HD)) into two non-overlapping sets: {\em Adv Set} $\mathcal{A}$ and {\em Detector Set} $\mathcal{D}$ (see Table~\ref{dataset_Split}). $\mathcal{A}$ and $\mathcal{D}$ each contain half of the instances of each class. This means, in both $\mathcal{A}$ and $\mathcal{D}$, 259 instances for each of 95 classes for FD data and 360 instances for each of 83 classes for HD data. 
%Note that this means our attacker models will have slightly lower accuracies than the results reported in prior work on DL attacks.
%For FD data, each set contains 259 instances from each of 95 classes, while for HD data, each set contains 83 classes each with 360 instances.

%% file: figures/both_cdf_bursts.tex
% This file was created by matplotlib2tikz v0.5.7.
% The lastest updates can be retrieved from
% 
% https://github.com/nschloe/matplotlib2tikz
% 
% where you can also submit bug reports and leavecomments.
% 
\begin{tikzpicture}
\begin{axis}[
xlabel={Number of Bursts},
ylabel={Cumulative Fraction},
xmin=0, xmax=1500,
ymin=0, ymax=1,
axis on top,
xmajorgrids,
ymajorgrids,
grid=both,
xtick distance= 250,
width=0.950*\figurewidth,
height=.750*\figureheight,
%width=0.90*\figurewidth,
%height=.90*\figureheight,
tick pos=left,
xmajorgrids,
ymajorgrids,
style={mark size=2pt},
x grid style={lightgray!115!black},
y grid style={lightgray!115!black},
legend style={draw=white!120.0!black, scale=.9, legend columns=1, at={(0.45,0.4)}, anchor=west, font=\small},
legend entries={{Half-Duplex},{Full-Duplex}},
legend cell align={right}
]
\addplot [line width = 0.5mm,red!50.0!black]
table {%
19 0.0118911919460884
19 0.0171255049653022
25 0.059455959730442
28 0.0877431356635449
29 0.0900320077329224
29 0.0907068146378384
31 0.100345610563464
39 0.127483790955764
39 0.128696619582167
39 0.129034023034625
77 0.145840362571926
103 0.174647322202059
105 0.177291835748352
111 0.195301884899828
113 0.200016414222011
119 0.215847019450853
119 0.218692151266175
120 0.219157220889833
129 0.237221984114681
130 0.238799573230228
139 0.251073763689917
141 0.253234969588094
145 0.259791539380454
151 0.266731107686415
157 0.275375931279124
157 0.27669818805227
179 0.312189383645963
179 0.313584592516939
187 0.326323852600286
191 0.333801442627735
191 0.334102370031278
203 0.356507783076937
203 0.356790472456024
215 0.378867601061453
219 0.385515360976099
228 0.396923245274072
233 0.403206244699574
242 0.412735612478456
250 0.421097746692078
267 0.43959110349167
267 0.440311505457729
271 0.445481985391342
279 0.456233300808856
283 0.463382606396075
295 0.481255870364122
314 0.513126818103063
323 0.525911673247554
341 0.552976901542025
345 0.557846454072095
347 0.562196222905135
357 0.576567786177401
357 0.577342902216832
367 0.590829921302924
367 0.591741822525784
373 0.599967171555977
389 0.619135335260485
404 0.637382478729904
406 0.639589279689224
421 0.657125140204813
430 0.666517722800266
431 0.668077073891356
433 0.669882638312618
436 0.672746008152397
437 0.673867646656514
438 0.674651881708173
443 0.678937817455613
465 0.694959921941255
516 0.723064717629786
527 0.729010313602831
533 0.732721751579869
543 0.737819279415654
561 0.750193779009858
571 0.756695634728846
579 0.762723301811948
585 0.766352668678929
592 0.771176626147856
596 0.773757306608548
609 0.782137678846627
611 0.783861172157832
623 0.792013569090196
632 0.798725162090442
637 0.802327171920738
641 0.805692087433089
647 0.809221145165556
657 0.816224546557117
658 0.816808163339747
689 0.836614657900256
695 0.839660407984607
702 0.843727487438561
719 0.85366721076773
727 0.858682667493457
827 0.909329661411076
833 0.911399677186967
857 0.921084068173735
870 0.926692260694322
880 0.931242647796391
953 0.951295355687072
1176 0.976390877340166
1330 0.985601079691048
1411 0.989339874704772
};
\addplot [line width = 0.5mm,green!50.0!black, dashed]
table {%
10.0 0.11169452510268246
68.05 0.25851926519927165
126.1 0.38798200610829414
184.14999999999998 0.49090526125746575
242.2 0.5817473332631227
300.25 0.6448312696526091
358.29999999999995 0.7021529480794976
416.34999999999997 0.7661245430063033
474.4 0.8199256774038238
532.4499999999999 0.8549054417981848
590.5 0.879428889524124
648.55 0.8983254848270563
706.5999999999999 0.9139573021198482
764.65 0.9255570433448169
822.6999999999999 0.9341628176388275
880.75 0.9409330946183813
938.8 0.9469962537800706
996.8499999999999 0.9528939172822597
1054.8999999999999 0.9573171649089013
1112.95 0.961228880497088
1171.0 0.9641476221282734
1229.05 0.9665548317210038
1287.1 0.9687965456542337
1345.1499999999999 0.9709780793476454
1403.1999999999998 0.9735507845998758
1461.25 0.975717273233333
1519.3 0.9777784464471082
1577.35 0.979373222802292
1635.3999999999999 0.9805016022988843
1693.4499999999998 0.9814795311959309
1751.5 0.9823220545533865
1809.55 0.9829689921314327
1867.6 0.9834504340499788
1925.6499999999999 0.9840672815081158
1983.6999999999998 0.9846991740262077
2041.75 0.9852558412445266
2099.7999999999997 0.9859629590623912
2157.85 0.9867302571200739
2215.9 0.987587825537484
2273.95 0.9885055741947124
2332.0 0.989483503091759
2390.0499999999997 0.990070260429987
2448.1 0.9906570177682151
2506.15 0.9910481893270336
2564.2 0.9916650367851708
2622.25 0.9921464787037169
2680.2999999999997 0.9925075601426264
2738.35 0.9930792724208998
2796.3999999999996 0.993485489039673
2854.45 0.993831525418628
2912.5 0.9941625167376283
2970.5499999999997 0.9943881926369468
3028.6 0.9945837784163563
3086.6499999999996 0.994899724675402
3144.7 0.9952156709344477
3202.75 0.9955767523733573
3260.7999999999997 0.9958024282726758
3318.85 0.9961183745317216
3376.8999999999996 0.9962688251312672
3434.95 0.9964644109106766
3493.0 0.9966900868099952
3551.0499999999997 0.9967954022296771
3609.1 0.9969458528292228
3667.1499999999996 0.9972016188484504
3725.2 0.9973671145079505
3783.25 0.9974272947477686
3841.2999999999997 0.9975777453473142
3899.35 0.9977582860667691
3957.3999999999996 0.9978936916063601
4015.45 0.9980140520859967
4073.5 0.9981494576255877
4131.55 0.9982698181052242
4189.599999999999 0.9983450434049971
4247.65 0.9984654038846337
4305.7 0.9986008094242247
4363.75 0.9986910797839521
4421.8 0.9987813501436796
4479.849999999999 0.9989318007432253
4537.9 0.999082251342771
4595.95 0.9992778371221803
4654.0 0.999428287721726
4712.05 0.9995486482013626
4770.099999999999 0.9995787383212716
4828.15 0.9996690086809992
4886.2 0.9996990988009082
4944.25 0.9997893691606358
5002.3 0.9998796395203632
5060.349999999999 0.9998946845803178
5118.4 0.9999097296402725
5176.45 0.9999247747002271
5234.5 0.9999247747002271
5292.55 0.9999247747002271
5350.599999999999 0.9999247747002271
5408.65 0.9999247747002271
5466.7 0.9999247747002271
5524.75 0.9999548648201363
5582.799999999999 0.9999548648201363
5640.849999999999 0.9999548648201363
5698.9 0.9999849549400454
5756.95 1.0
};
\end{axis}
\end{tikzpicture}

%% file: tables/dataset_split.tex
\begin{table}[tbp]
\centering
\renewcommand{\arraystretch}{1.1}
%\label{dataset_Split}
\caption{Dataset Split: {\em Adv Set} ($\mathcal{A}$)  \& {\em Detector Set ($\mathcal{D}$)}. {\bf FD}: Full-Duplex, {\bf HD}: Half-Duplex, {\bf C}: Class, {\bf I}: Instance, {\bf CW}:Closed-World, {\bf OW}: Open-World.}
\begin{tabular}{lcccc} 
%\label{dataset_Split}
%\hline
    & {\em Adv Set} $\mathcal{A}$  & {\em Detector Set} $\mathcal{D}$ & CW   & OW  \\
    & ($C \times I$) & ($C\times I$) & Total   &   \\ 
\hline
\rule{0pt}{2ex}
FD         & 95$\times$259           & 95$\times$259      & 95$\times$518 & 40,716   \\
~HD & 83$\times$360       & 83$\times$360       & 83$\times$720  & 40,000   \\
%\hline
\label{dataset_Split}     
%\vspace{-0.6cm}
\end{tabular}
\end{table}

%% file: figures/alpha.tex
\begin{tikzpicture}

\begin{axis}[%
grid=major,grid style={dashed},
width=1in,
height=1in,
scale only axis,
xmin=0,
xmax=8,
separate axis lines,
every outer y axis line/.append style={black},
legend style={draw=white!80.0!black, scale=0.2, legend columns=1, at={(0.175,0.90)}, anchor=west, font=\tiny},
legend entries={\ref{bw_case_1}{Case I},\ref{bw_case_2}{Case II}},
legend cell align={left},
every y tick label/.append style={font=\color{black}},
ymin=0.25,
ymax=0.75,
ytick       ={0, 0.25, 0.35,0.45,0.55,  0.65,0.75,    1},
ylabel      ={\ref{acc_case_1_}\ref{acc_case_2_} Accuracy}, %\ref{acc_case_1_}\ref{acc_case_2_}
yticklabels ={0, 0.25, 0.35,0.45,0.55,  0.65,0.75,    1},
xtick       ={1, 3, 5, 7},
xlabel      ={$\alpha$ Value},
%xticklabels ={0.00,,,,,0.005,,,,,0.010,,,,,0.015,,,,,0.020},
%xtick      ={0,2.5,5,7.5,10,12.5,15,17.5,20,21},
%xticklabels    ={0.00,,0.005,,0.010,,0.015,,0.020,},
]
\addplot [
color=red,
solid,
line width=2.0pt,
mark = *,
]
table[row sep=crcr]{
1	0.453880537\\
3	0.393742381\\
5	0.352295815\\
7	0.351076798\\
}; \label{acc_case_1_}

\addplot [
color=red,
dashed,
line width=2.0pt,
mark = *,
]
table[row sep=crcr]{
1	0.629418935\\
3	0.577813896984\\
5	0.555465258\\
7	0.564811053\\
}; \label{acc_case_2_}
\end{axis}

\begin{axis}[%
width=1in,
height=1in,
scale only axis,
xmin=0,
xmax=8,
every outer y axis line/.append style={black},
every y tick label/.append style={font=\color{black}},
ymin=0.25,
ymax=0.75,
ytick={0, 0.25, 0.35,0.45,0.55,  0.65,0.75,    1},
ylabel={\ref{bw_case_1_}\ref{bw_case_2_} BWO}, %\ref{bw_case_1_}\ref{bw_case_2_} 
axis x line*=bottom,
axis y line*=right,
xtick={},
xticklabels={},
]
\addplot [
color=blue,
solid,
line width=2.0pt,
mark = +,
forget plot
]
table[row sep=crcr]{
1	0.398683953\\
3	0.519205474\\
5	0.56509686\\
7	0.584732025\\
}; \label{bw_case_1_}

\addplot [
color=blue,
dashed,
line width=2.0pt,
mark = +,
forget plot
]
table[row sep=crcr]{
1	0.37785661\\
3   0.509568787\\
5   0.565623718\\
7   0.58487858\\
}; \label{bw_case_2_}
\end{axis}
\end{tikzpicture}%

%% file: figures/C_1_full_acc_bw.tex
\begin{tikzpicture}

\begin{axis}[%
grid=major,grid style={dashed},
width=1in,
height=1in,
scale only axis,
xmin=100,
xmax=500,
separate axis lines,
every outer y axis line/.append style={black},
legend style={draw=white!80.0!black, scale=0.2, legend columns=1, at={(0.05,0.8750)}, anchor=west, font=\tiny},
legend entries={\ref{bw_1_}{$\alpha$=5},\ref{bw_2_}{$\alpha$=7}},
legend cell align={left},
every y tick label/.append style={font=\color{black}},
ymin=0.25,
ymax=0.75,
ytick       ={0, 0.25, 0.35,0.45,0.55,  0.65,0.75,    1},
ylabel      ={\ref{acc_1_}\ref{acc_2_} Accuracy},
yticklabels ={0, 0.25, 0.35,0.45,0.55,  0.65,0.75,    1},
xtick       ={100,200, 300,400, 500},
xlabel      ={Number of Iterations},
%xticklabels ={0.00,,,,,0.005,,,,,0.010,,,,,0.015,,,,,0.020},
%xtick      ={0,2.5,5,7.5,10,12.5,15,17.5,20,21},
%xticklabels    ={0.00,,0.005,,0.010,,0.015,,0.020,},
]
\addplot [
color=red,
solid,
line width=2.0pt,
mark = *,
]
table[row sep=crcr]{
100	0.488013003\\
200	0.427468509\\
300	0.383990248\\
400	0.357171881\\
500	0.352295815\\
}; \label{acc_1_}

\addplot [
color=red,
dashed,
line width=2.0pt,
mark = *,
]
table[row sep=crcr]{
100	0.416497359\\
200	0.392117026\\
300	0.38317757\\
400	0.380333198\\
500	0.351076798\\
}; \label{acc_2_}
\end{axis}

\begin{axis}[%
width=1in,
height=1in,
scale only axis,
xmin=100,
xmax=500,
every outer y axis line/.append style={black},
every y tick label/.append style={font=\color{black}},
ymin=0.25,
ymax=0.75,
ytick={0, 0.25, 0.35,0.45,0.55,  0.65,0.75,    1},
ylabel={\ref{bw_1_}\ref{bw_2_}BWO},
axis x line*=bottom,
axis y line*=right,
xtick={},
xticklabels={},
]
\addplot [
color=blue,
solid,
line width=2.0pt,
mark = +,
forget plot
]
table[row sep=crcr]{
100	0.341516136\\
200	0.447646906\\
300	0.507402166\\
400	0.540677202\\
500	0.56509686\\
}; \label{bw_1_}

\addplot [
color=blue,
dashed,
line width=2.0pt,
mark = +,
forget plot
]
table[row sep=crcr]{
100	0.388124251\\
200	0.488895193\\
300	0.534805105\\
400	0.569287315\\
500	0.584732025\\
}; \label{bw_2_}
\end{axis}
\end{tikzpicture}%

%% file: figures/C_2_full_acc_bw.tex
\begin{tikzpicture}

\begin{axis}[%
grid=major,grid style={dashed},
width=1in,
height=1in,
scale only axis,
xmin=100,
xmax=500,
separate axis lines,
every outer y axis line/.append style={black},
legend style={draw=white!80.0!black, scale=0.2, legend columns=1, at={(0.20,0.90)}, anchor=west, font=\tiny},
legend entries={\ref{bw_1}{$\alpha$=5},\ref{bw_2}{$\alpha$=7}},
legend cell align={left},
every y tick label/.append style={font=\color{black}},
ymin=0.25,
ymax=0.75,
ytick       ={0, 0.25, 0.35,0.45,0.55,  0.65,0.75,    1},
ylabel      ={\ref{accr_1}\ref{accr_2} Accuracy},
yticklabels ={0, 0.25, 0.35,0.45,0.55,  0.65,0.75,    1},
xtick       ={100,200, 300,400, 500},
xlabel      ={Number of Iterations},
%xticklabels ={0.00,,,,,0.005,,,,,0.010,,,,,0.015,,,,,0.020},
%xtick      ={0,2.5,5,7.5,10,12.5,15,17.5,20,21},
%xticklabels    ={0.00,,0.005,,0.010,,0.015,,0.020,},
]
\addplot [
color=red,
solid,
line width=2.0pt,
mark = *,
]
table[row sep=crcr]{
100	0.672490857\\
200	0.619666802\\
300	0.568061763\\
400	0.576188541\\
500	0.555465258\\
}; \label{accr_1}

\addplot [
color=red,
dashed,
line width=2.0pt,
mark = *,
]
table[row sep=crcr]{
100	0.668427469\\
200	0.607476635\\
300	0.55465258\\
400	0.562373019\\
500	0.564811053\\
}; \label{accr_2}
\end{axis}

\begin{axis}[%
width=1in,
height=1in,
scale only axis,
xmin=100,
xmax=500,
every outer y axis line/.append style={black},
every y tick label/.append style={font=\color{black}},
ymin=0.25,
ymax=0.75,
ytick={0, 0.25, 0.35,0.45,0.55,  0.65,0.75,    1},
ylabel={\ref{bw_1}\ref{bw_2}BWO},
axis x line*=bottom,
axis y line*=right,
xtick={},
xticklabels={},
]
\addplot [
color=blue,
solid,
line width=2.0pt,
mark = +,
forget plot
]
table[row sep=crcr]{
100	0.321277019\\
200	0.428315128\\
300	0.486737203\\
400	0.537375876\\
500	0.565623718\\
}; \label{bw_1}

\addplot [
color=blue,
dashed,
line width=2.0pt,
mark = +,
forget plot
]
table[row sep=crcr]{
100	0.365710726\\
200	0.467772658\\
300	0.5301908\\
400	0.564467784\\
500	0.58487858\\
}; \label{bw_2}
\end{axis}
\end{tikzpicture}%

%% file: tables/hyper_para.tex
\begin{table}[t!]
\centering
\renewcommand{\arraystretch}{1.1}
\caption{Hyperparameter Tuning on DF and Var-CNN attack models for Black-box attacks.}
\begin{tabular}{lccc} 
%\hline
\multirow[b]{2}{*}{\textbf{Hyperparameters}} & \multirow[b]{2}{*}{\textbf{Choices}}  & \multicolumn{2}{c}{\textbf{Final}} \\ \cline{3-4}
 &  & \textbf{DF} & \textbf{Var-CNN}  \\
\hline
\rule{0pt}{2ex}
Training Epoch & $[30, 50, 100, 150]$ & 100 & 100 \\
~Batch Size & $[32, 64, 128, 256]$ & 32 & 32\\
~Optimizer & $[$Adam, Adamax$]$ & Adamax & Adamax\\
~Learning Rate & $[0.001, 0.002, 0.003]$  & 0.002 & 0.002 \\
~Activation Fn. & $[$ReLU, ELU$]$ & ReLU, ELU & ReLU \\
%\hline
\end{tabular}
%\vspace{-0.3cm}
\label{hyper_para}
\end{table}

%% file: figures/C1_half_acc_bw.tex
\begin{tikzpicture}

\begin{axis}[%
grid=major,grid style={dashed},
width=1in,
height=1in,
scale only axis,
xmin=1,
xmax=7,
separate axis lines,
every outer y axis line/.append style={black},
legend style={draw=white!80.0!black, scale=0.2, legend columns=1, at={(-1.65,0.65)}, anchor=west, font=\tiny},
legend entries={\ref{bw_case_1}{Case I},\ref{bw_case_2}{Case II}},
legend cell align={left},
every y tick label/.append style={font=\color{black}},
ymin=0.25,
ymax=0.75,
ytick       ={0, 0.25, 0.35,0.45,0.55,  0.65,0.75,    1},
ylabel      ={\ref{acc_case_1}\ref{acc_case_2} Accuracy},
yticklabels ={0, 0.25, 0.35,0.45,0.55,  0.65,0.75,    1},
xtick       ={1, 3, 5, 7},
xlabel      ={$\alpha$ Value},
%xticklabels ={0.00,,,,,0.005,,,,,0.010,,,,,0.015,,,,,0.020},
%xtick      ={0,2.5,5,7.5,10,12.5,15,17.5,20,21},
%xticklabels    ={0.00,,0.005,,0.010,,0.015,,0.020,},
]
\addplot [
color=red,
solid,
line width=2.0pt,
mark = *,
]
table[row sep=crcr]{
1	0.656626506\\
3	0.581994645\\
5	0.508032128\\
7	0.355421686787\\
}; \label{acc_case_1}

\addplot [
color=red,
dashed,
line width=2.0pt,
mark = +,
]
table[row sep=crcr]{
1	0.596385542\\
3	0.539156626\\
5	0.500669344\\
7	0.288152610462\\
}; \label{acc_case_2}
\end{axis}

\begin{axis}[%
width=1in,
height=1in,
scale only axis,
xmin=1,
xmax=7,
every outer y axis line/.append style={black},
every y tick label/.append style={font=\color{black}},
ymin=0.25,
ymax=0.75,
ytick={0, 0.25, 0.35,0.45,0.55,  0.65,0.75,    1},
ylabel={\ref{bw_case_1}\ref{bw_case_2} BWO},
axis x line*=bottom,
axis y line*=right,
xtick={},
xticklabels={},
]
\addplot [
color=blue,
solid,
line width=2.0pt,
mark = *,
forget plot
]
table[row sep=crcr]{
1	0.465368793\\
3	0.511928553\\
5	0.577861167\\
7	0.627250120499\\
}; \label{bw_case_1}

\addplot [
color=blue,
dashed,
line width=2.0pt,
mark = +,
forget plot
]
table[row sep=crcr]{
1	0.569562178\\
3   0.614832145\\
5   0.696390129\\
7   0.734958089285\\
}; \label{bw_case_2}
\end{axis}
\end{tikzpicture}%

% Case 1: a: 355421686787, b: 0.627250120499
% Case 2: a: 288152610462, b: 0.734958089285

%% file: figures/C1_half_its.tex
\begin{tikzpicture}

\begin{axis}[%
grid=major,grid style={dashed},
width=1in,
height=1in,
scale only axis,
xmin=100,
xmax=500,
separate axis lines,
every outer y axis line/.append style={black},
legend style={draw=white!80.0!black, scale=0.2, legend columns=1, at={(-1.65,0.65)}, anchor=west, font=\tiny},
legend entries={\ref{half_bw_1}{Case I},\ref{half_bw_2}{Case II}},
legend cell align={left},
every y tick label/.append style={font=\color{black}},
ymin=0.25,
ymax=0.75,
ytick       ={0, 0.25, 0.35,0.45,0.55,  0.65,0.75,    1},
ylabel      ={\ref{half_acc_1}\ref{half_acc_2} Accuracy},
yticklabels ={0, 0.25, 0.35,0.45,0.55,  0.65,0.75,    1},
xtick       ={100,200, 300,400, 500},
xlabel      ={Number of Iterations},
%xticklabels ={0.00,,,,,0.005,,,,,0.010,,,,,0.015,,,,,0.020},
%xtick      ={0,2.5,5,7.5,10,12.5,15,17.5,20,21},
%xticklabels    ={0.00,,0.005,,0.010,,0.015,,0.020,},
]
\addplot [
color=red,
thin,
line width=2.0pt,
mark = +,
]
table[row sep=crcr]{
100	0.505354752343\\
200	0.444779116506\\
300	0.395582329237\\
400	0.370816599772\\
500	0.355421686787\\
}; \label{half_acc_1}

\addplot [
color=red,
dashed,
line width=2.0pt,
mark = *,
]
table[row sep=crcr]{
100	0.453815261124\\
200	0.367469879558\\
300	0.326974564886\\
400	0.299196787109\\
500	0.288152610462\\
}; \label{half_acc_2}
\end{axis}

\begin{axis}[%
width=1in,
height=1in,
scale only axis,
xmin=100,
xmax=500,
every outer y axis line/.append style={black},
every y tick label/.append style={font=\color{black}},
ymin=0.25,
ymax=0.75,
ytick={0, 0.25, 0.35,0.45,0.55,  0.65,0.75,    1},
ylabel={\ref{half_bw_1}\ref{half_bw_2} BWO},
axis x line*=bottom,
axis y line*=right,
xtick={},
xticklabels={},
]
\addplot [
color=blue,
thin,
line width=2.0pt,
mark = +,
forget plot
]
table[row sep=crcr]{
100	0.336177145545\\
200	0.458365795974\\
300	0.532211035815\\
400	0.591905950057\\
500	0.627250120499\\
}; \label{half_bw_1}

\addplot [
color=blue,
dashed,
line width=2.0pt,
mark = *,
forget plot
]
table[row sep=crcr]{
100	0.47831918778\\
200	0.589031717287\\
300	0.654786353921\\
400	0.694351645984\\
500	0.734958089285\\
}; \label{half_bw_2}
\end{axis}
\end{tikzpicture}%

%% file: tables/results_analysis.tex
\begin{table}[t!]
\centering

\caption{\textbf{White-Box.} Evaluation of \system~against DF. %and AWF attacks. %, CUMUL, $k$-FP, and $k$-NN Attacks \& Comparison against WTF-PAD and W-T Defenses. \textbf{S$\times$I}: Sites$\times$Instances, 
\textbf{BWO}: Bandwidth Overhead, \textbf{FD}: Full-Duplex, \textbf{HD}: Half-Duplex.}
\begin{tabular}{ll|cc} 
\textbf{Cases} & \textbf{Dataset} & \textbf{BWO} & \textbf{DF~\cite{sirinamDF}}  \\% & \textbf{AWF~\cite{Rimmer2018}} \\% & \textbf{CUMUL~\cite{cumul}} & \textbf{$k$-FP}~\cite{kfp} & \textbf{$k$-NN}~\cite{Wang2014} & {\bf DF Top-2} \\ 
\hline
\rule{0pt}{2ex}
\multirow{2}{*}{\textbf{Case I}}& 
FD  & {0.56}      & {0.35} \\% & {0.20} \\%&  {0.232} & {0.21} & {0.04} & {\bf 0.51}\\
& HD  &   {0.63}     & {0.35} \\% & {0.27} \\ %&   {0.300} & 0.33 & {0.06} & {\bf 0.52}\\
%\hline%\hline

\rule{0pt}{3ex}
\multirow{2}{*}{\textbf{Case II}}& FD &{0.56}     &  0.55 \\ %& {0.34} \\ %& 0.36  & 0.34 & {0.06} & {\bf 0.56}\\
& HD  &   0.73     & {0.29} \\ %& {0.22} \\ %&  {0.31} & 0.30 & {0.05} & {\bf 0.43}\\
\end{tabular}
\label{tab_results_analysis}
\end{table}

%% file: tables/surrogate.tex
\begin{table*}[t!]
\centering
\renewcommand{\arraystretch}{1.1}
\caption{{\bf Black-Box.} %Evaluation of \system~when 
AWF CNN as Detector. Evaluation of \system~against DF, Var-CNN, CUMUL, $k$-FP, and $k$-NN attacks \& comparison against WTF-PAD and W-T defenses. \textbf{BWO}: Bandwidth Overhead, \textbf{FD}: Full-Duplex, \textbf{HD}: Half-Duplex.}
\label{tab:black-box}
\begin{tabular}{ll|c|c c|c c c | c c}
%\hline 
{\bf Case}          & {\bf Dataset} & {\bf BWO}    & {\bf DF~\cite{sirinamDF}} & {\bf Var-CNN~\cite{bhat2019var}}
 & \textbf{CUMUL~\cite{cumul}} & \textbf{$k$-FP}~\cite{kfp} & \textbf{$k$-NN}~\cite{Wang2014} & {\bf DF Top-2}  & {\bf Var-CNN Top-2}   \\ 
 \hline
& Undefended (FD)   &  - & 0.97 &  0.98 & 0.93 & 0.85 & 0.86 & - & -\\
& Undefended (HD)  &  - & 0.98 & 0.99 & 0.92 & 0.92 & 0.90 & - & -\\
& WTF-PAD~\cite{wtf-pad} & 0.64 & 0.86 & 0.90 & 0.55 & 0.44 & 0.17  & 0.92 & 0.95\\ 
& W-T~\cite{Wang2017} &  0.72 & 0.40 & 0.44 & 0.36 & 0.30 & 0.35 & 0.97 & 0.94\\ 

%\hline %\hline

\rule{0pt}{4ex}
\multirow{2}{*}{\textbf{Case I}} 
    & \system~(FD)     &  0.58     & 0.42 & 0.35 
    &   0.19    &   0.21    & 0.08      &  0.56 &  0.50 \\ 
    &  \system~(HD)  &    0.62   &   0.41  & 0.33 
    &   0.22    &   0.28    & 0.10    & 0.57  &  0.47  \\ 

\rule{0pt}{4ex}
\multirow{2}{*}{\textbf{Case II}} 
    &   \system~(FD)   &    0.58   & 0.58 & 0.62 
    &   0.32    &   0.32    & 0.14        &  0.70 &  0.72 \\
    & \system~(HD)     &  0.70     &  0.38 & 0.30 
    &   0.20    &   0.26    & 0.12         &  0.54 &  0.43 \\
% \hline
\end{tabular}
\label{weak_detector}
\vskip -0.3cm
\end{table*}

%% file: figures/top_k_all.tex
\begin{tikzpicture}

\begin{axis}[%
grid=major,grid style={dashed},
width=2.5in,
height=1in,
scale only axis,
xmin=1,
xmax=10,
separate axis lines,
every outer y axis line/.append style={black},
legend style={draw=white!80.0!black, scale=0.2, legend columns=2, at={(0.30,0.20)}, anchor=west, font=\small},
legend entries={{FD Case I},{FD Case II},{HD Case I},{HD Case II}},
legend cell align={left},
every y tick label/.append style={font=\color{black}},
ymin=0.25,
ymax=0.95,
ytick       ={0, 0.25, 0.35,0.45,0.55,  0.65,0.75, 0.85,0.95,1},
ylabel      ={Accuracy},
yticklabels ={0, 0.25, 0.35,0.45,0.55,  0.65,0.75, 0.85, 0.95,   1},
xtick       ={1,2, 3,4, 5,6,7,8,9,10},
xlabel      ={Top-$k$},
%xticklabels ={0.00,,,,,0.005,,,,,0.010,,,,,0.015,,,,,0.020},
%xtick      ={0,2.5,5,7.5,10,12.5,15,17.5,20,21},
%xticklabels    ={0.00,,0.005,,0.010,,0.015,,0.020,},
]
\addplot [
color=black,
thin,
line width=2.0pt,
mark = *,
]
% FD Case I: [0.3823648989200592, 0.5323039293289185, 0.627793550491333, 0.6980901956558228, 0.7431938052177429, 0.7854530811309814, 0.8138967752456665, 0.8342137336730957, 0.8590003848075867, 0.8772856593132019]
%tuned DF [0.4205607476393317, 0.5774075580486094, 0.659488012881746, 0.7171881350013891, 0.7631044294087201, 0.7943925236793419, 0.8155221457841774, 0.8346200734558555, 0.8545306784164028, 0.8679398618932179]
table[row sep=crcr]{
1	0.4205607476393317\\
2	0.5774075580486094\\
3	0.659488012881746\\
4	0.7171881350013891\\
5	0.7631044294087201\\
6   0.7943925236793419 \\
7   0.8155221457841774 \\
8   0.8346200734558555 \\
9   0.8545306784164028 \\
10  0.8679398618932179 \\
}; \label{fd_c1_k}
\addplot [
color=black,
dashed,
line width=2.0pt,
mark = *,
]
% FD Case II: [0.5103616118431091, 0.6692401170730591, 0.7517269253730774, 0.7927671670913696, 0.837464451789856, 0.8606257438659668, 0.8821617364883423, 0.9000406265258789, 0.9126371145248413, 0.9232019782066345]
% Tuned DF [0.5821308407824197, 0.7293783016003326, 0.7972368956193455, 0.8370581062428112, 0.8691588782866957, 0.8854124337519538, 0.8984152781241602, 0.908167411403315, 0.9175132057958384, 0.9264526613017303]
table[row sep=crcr]{
1	0.5821308407824197\\
2	0.7293783016003326\\
3	0.7972368956193455\\
4	0.8370581062428112\\
5	0.8691588782866957\\
6   0.8854124337519538 \\
7   0.8984152781241602 \\
8   0.908167411403315 \\
9   0.9175132057958384 \\
10  0.9264526613017303 \\
}; \label{fd_c2_k}
\addplot [
color=red,
thin,
line width=2.0pt,
mark = +,
]
% HD Case I: [0.40662649273872375, 0.5672690868377686, 0.6646586060523987, 0.7329317331314087, 0.7881526350975037, 0.8263052105903625, 0.8490629196166992, 0.8711512684822083, 0.8882195353507996, 0.9019411206245422]
table[row sep=crcr]{
1	0.40662649273872375\\
2	0.5672690868377686\\
3	0.6646586060523987\\
4	0.7329317331314087\\
5	0.7881526350975037\\
6   0.8263052105903625 \\
7   0.8490629196166992 \\
8   0.8711512684822083 \\
9   0.8882195353507996 \\
10  0.9019411206245422 \\
}; \label{hd_c1_k}
\addplot [
color=red,
dashed,
line width=2.0pt,
mark = +,
]
% HD Case II: [0.3755020201206207, 0.5361445546150208, 0.6358768343925476, 0.6984605193138123, 0.7503346800804138, 0.7955154180526733, 0.8219544887542725, 0.8460508584976196, 0.8674699068069458, 0.8818607926368713]
table[row sep=crcr]{
1	0.3755020201206207\\
2	0.5361445546150208\\
3	0.6358768343925476\\
4	0.6984605193138123\\
5	0.7503346800804138\\
6   0.7955154180526733 \\
7   0.8219544887542725 \\
8   0.8460508584976196 \\
9   0.8674699068069458 \\
10  0.8818607926368713 \\
}; \label{hd_c2_k}
\end{axis}
\end{tikzpicture}%

%% file: tables/multi_round.tex
\begin{table}[t!]
\centering
\renewcommand{\arraystretch}{1.1}
\caption{\textbf{Intersection Attack.} Evaluation of \system~against an intersection attack over five rounds.
\textbf{FD}: Full-Duplex, \textbf{HD}: Half-Duplex.}
\begin{tabular}{ll|ccc} 
%\hline 
\multirow[b]{2}{*}{\textbf{Cases}} & \multirow[b]{2}{*}{\textbf{Dataset}} & \textbf{Absolute} & \textbf{Absolute} & \textbf{Mean} \\
& & \textbf{Success} & \textbf{Failure} & \textbf{Intersection} \\
\hline%\hline
\multirow{2}{*}{~\textbf{I}}& 
\system~(FD)  & {0.27}     & {0.51} & 2.86\\
& \system~(HD)  &   {0.17}     & {0.52} & 4.12\\%

\rule{0pt}{4ex}
\multirow{2}{*}{\textbf{II}}& \system~(FD) &{0.36}     &  {0.20} & 2.60\\
& \system~(HD)  &   {0.13}     & {0.53} & 4.12\\
%\hline
\end{tabular}
\label{multiround}
\end{table}

%% file: 6_discussion.tex
\section{Discussion}
\label{diss}
%\subsection{Difference between W-T \& \system}
%\label{diff_wt_mock}

% SLIDE paper citation - \cite{chen2019slide}

\subsection{Comparison}\label{comparison}

%\todo{Update for Var-CNN result comparison}
\system~has significant advantages over the state-of-the-art defenses. Compared to WTF-PAD, the Top-1 accuracies of DF and Var-CNN are at least 28\% lower, effectively tripling the error rate for the attacker. Compared to W-T, the Top-1 accuracies of DF and Var-CNN are relatively higher, but the Top-2 accuracies are at least 27\% and 22\% lower, respectively. We believe that the Top-2 accuracy of 97\% for W-T should be considered unacceptable given the costs of deploying and running the defense.

One question that our findings raise is \emph{Why does \system~perform so well in Top-$k$ accuracy?} We hypothesize that the random selection of targets during the search for an adversarial example is helping to create unpredictable patterns that not only move away from the original site class but move towards a number of different possible classes. W-T is explicitly designed to provide confusion only among a limited number of sites, two by default. WTF-PAD has random patterns, but these may only lead to confusion among sites that are already similar. \system~can find new sites that confuse the attacker.

Note that our black-box models assume that we use a detector model (AWF) that is weaker than state-of-the-art models (DF and Var-CNN). Employing a more powerful detector model could thus have the potential to make \system~robust against even more powerful models that may be developed in the future. 

Unlike W-T, \system~can be deployed with full-duplex communication and thus lower latency overhead~\cite{Wang2017}. Finally, \system's bandwidth overhead in full-duplex mode is modestly better than both of these defenses.

\subsection{Implementation \& Deployment}\label{deployment_plan}

% model compression & other techniques to improve runtime performance
% adversarial patches
To deploy \system~in the real world, it is necessary to address several outstanding issues.

\paragraphX{\textbf{Live Trace Generation.}} We have yet to find a solution that allows for live packet-by-packet generation of adversarial traces. Consequently, \system~requires that the full traffic burst sequence be known before generating an associated adversarial example. This requirement is also in the Walkie-Talkie defense~\cite{Wang2017}, along with several more expensive defenses. It means that the defense must maintain a database of relatively recent reference burst sequences for sites of interest. The responsibility for gathering and maintaining this database would most appropriately be given to the Tor network. In particular, special directory servers could be deployed to maintain the database and distribute fresh reference sequences periodically to clients. The servers could also be used to distribute pre-trained detector models to clients. The clients can then use this data to generate adversarial traces locally before each site visit.

Research on \emph{adversarial patches} shows that adversarial examples can be generated that are agnostic to a particular target distribution and yet cause the source class to be misclassified by the model~\cite{moosavi2017universal, brown2017adversarial}. We leave investigating applying this idea to network traffic traces as potential future work.

\paragraphX{\textbf{Padding Mechanism.}} Padding mechanisms must be designed so they can manipulate a trace sequence to match an adversarial sequence. To address this problem, \emph{burst molding} must be performed by both the client and bridge, similar to that of the W-T defense~\cite{Wang2017}. Burst molding is difficult to achieve in a live setting, as padding must be added to the ends of bursts, which are difficult to reliably identify. To address this, we propose that a {burst molding} mechanism hold packets in a queue for each burst. The burst of real packets is forwarded normally until no additional packets are received for transmission. After a timeout is triggered, the queue is dumped onto the network. The size of the next burst can be easily communicated from the client to the bridge by embedding this information in the dummy packets of the prior burst. In this way, bursts can be easily molded to their appropriate size at the cost of a small additional delay. We leave the engineering challenge of setting the timeout for minimal added delay to future work.

\begin{figure}[!t]
    \centering
  \includegraphics[scale=0.45]{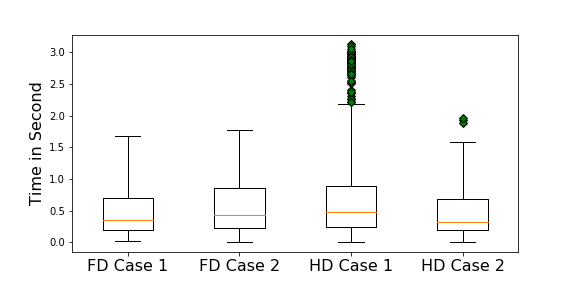}
  \caption{Time to generate adversarial traces in \system.}
  \label{fig:mb_gen_time}
  \vskip -0.5cm
\end{figure}

\paragraphX{\textbf{Computational Requirements.}} 
%Finally, there exists one more major issue impeding the deploy-ability of \system. 
The adversarial trace generation process currently requires hundreds of iterations of perturbing the reference trace and checking the resulting trace against the detector. For \system~to be deployable, the full generation process must take at most a few seconds before each visit. We can see the generation time of FD and HD data by \system~from Figure~\ref{fig:mb_gen_time}. For most of the sites, \system~can generate the trace in around 0.5 second. Some sites in HD data, however, take up to 3 seconds.

As previously noted, significant computational resources (e.g. a dedicated GPU device) are currently necessary to generate the trace within this time limitation. Fortunately, there are possible solutions to this problem. %At the moment, our algorithm performs adversarial trace generation using the state-of-the-art classifiers. These classifiers were designed to be used by comparatively high resource attackers who have access to GPU arrays that a client would likely not. This highlights a need for an alternative model that is more efficient than its state-of-the-art brethren. 

First, to save computation on the first visit to a site, we could relax the bandwidth constraint and enable the client to select a significantly different traffic pattern from the original with a single step. For sites that the user has visited previously, it would be possible to compute additional trace patterns in the background and save them for future use. 

Second, works in other domains have explored techniques that allow models to be run efficiently by low-resource entities such as mobile devices. Techniques such as model pruning~\cite{han2016-deepcompression, yu2018nisp} and data quantization~\cite{Zhu2017-ternararyquantization, jacob2018quantization} have been shown to provide significant speedups in resource-constrained environments without requiring extensive changes to the underlying model. In particular, Yu et al. ~\cite{yu2018nisp} demonstrate network quantization can achieve greater than 100\% speedup with only 2\% accuracy degredation on a MobileNet classifier. Similarly, Jacob et al.~\cite{jacob2018quantization} achieved up to a 67\% reduction in model computations using their parameter pruning technique on an AlexNet classifier. Additionally, techniques such as knowledge distillation~\cite{zagoruyko2016paying} can be used to train new, more efficient student models with minimal reductions in accuracy. Application of these techniques may achieve speedups in the range of 100-300\% for our detector model. %These techniques can be leveraged to create efficient detector models that consumer-grade devices can easily run to generate traces.

Finally, recent research by Chen et al.~\cite{chen2019slide} proposes a {\em Sub-LInear Deep learning Engine} (SLIDE) which enable a deep-learning model to run in a CPU even faster than a GPU, indicating that it is possible to deploy \system~in a realistic setting without requiring a dedicated GPU. 

We leave further investigation of these techniques to future work.

\subsection{Server-Side Defense}

Cherubin et al.~\cite{Cherubin2017} proposed to apply website fingerprinting defenses at the application layer, which is especially beneficial to onion services that are accessed only through Tor. A server-side application of \system~could be an effective and practical way to defend a site from attacks. It would involve the website operator running a tool to examine their site trace from Tor, running \system~to generate a set of effective traces, and then adding dummy web objects and padding existing web objects to specific sizes (following the techniques of Cherubin et al.~\cite{Cherubin2017}) to modify the network trace as needed. A web server can generate traces as needed during down times. Also, depending on the threat model of the site, it may only need to create a new trace periodically, such as once per day or per hour. At that rate of change, the attacker would need to download a new set of traces very frequently, increasing the attack cost and potentially making the attack less stealthy. %for less popular onion sites.

%% file: 7_Con_Future.tex
\section{Conclusion}
\label{conclusions-wfd}

We propose \system, a WF defense that 
%can significantly reduce the accuracy of state-of-the-art WF attacks using our technique for generating adversarial traces. \system~
offers better protection and lower bandwidth overhead than WTF-PAD and Walkie-Talkie, the previous state-of-the-art lightweight defenses. \system~uses a novel mechanism to create adversarial traces that are robust even against adversarial training. It drops the Top-1 accuracy of the best attack from over 98\% to at most 62\% and cuts the Top-2 accuracy to at most 72\%, which is much better than the Top-2 accuracies against WTF-PAD and Walkie-Talkie (95\% and 97\%, respectively). Furthermore, \system's full-duplex bandwidth overhead is 58\%, which is lower than either of the prior defenses.
%. It generates adversarial traces that can significantly limit the ability of a WF adversary to distinguish a site, even though the adversary adopts adversarial training to train his classifier. 
%Our defense mechanism results in 58\% bandwidth overhead and drops the accuracy of the state-of-the-art WF attack from 98\% to 23\%-56\%, depending on the scenario. In addition, our defense performs better than the two state-of-the-art defenses: WTF-PAD and W-T. The Top-2 attack accuracy of our defense against the DF attack is at most 67\%, whereas it is 90\% for WTF-PAD and 97\% for W-T. %To gain a more objective perspective on the quality of our defense, we analyze the information leakage of website features.
The results of information leakage analysis are in line with our previous conclusions when evaluating raw accuracy metrics. We emphasize that our experiments are conducted in the closed-world setting, where the attacker knows that the Tor client is assumed to visit one of the monitored sites. In a more realistic {\em open-world} setting, where the client could visit any site on the Web, 58\% accuracy is very likely to lead to many false positives for the attacker. 

Although \system~has implementation challenges that must be addressed before it could be practically deployed in Tor, it shows the significant potential of an approach inspired by adversarial examples. Furthermore, it may be possible to leverage \system~for server-side defense in the near future.

%Overall, our findings indicate that \system~can be a potential defense for Tor against WF attacks.

%% file: 8_Acks.tex
%\subsection*{Acknowledgement}
%This material is based upon work supported by the National Science Foundation under Grants No. 1423163, 1722743, and 1816851.

%This material is based upon work supported by the National Science Foundation under Awards No. 1722743, 1816851, and 1433736.

\paragraphX{\textbf{Acknowledgements.}} We thank all the anonymous reviewers for their valuable reviews and feedback in this work. This material is based upon work supported by the National Science Foundation (NSF) under Grants No. 1423163, 1722743, and 1816851.

%\balance

%% file: appen_A.tex
\section{Performance of Base C\&W Method}
\label{failed_carlini}

We evaluate the efficacy of the adversarial traces generated by the Carlini and Wagner~\cite{carlini2017towards} in this Section.

Given a sample $x$ and a model $F$, the algorithm finds a perturbation $\delta$ that makes $x' = x + \delta$ to be misclassified to any other class than $C(x)$ ($C(x) = {argmax}_{i}{F(x)}_{i}$), or in the case of a targeted attack, it is classified to target class $t$. The algorithm tries to find $x'$ that is similar to $x$ based on distance metric $D$. The distance metrics can be an ${L}_{p}$-norm such as ${L}_{0}$, ${L}_{2}$, or ${L}_{\infty}$.  The algorithm is formulated as follows:
\begin{IEEEeqnarray*}{rCl}
\textrm{min} \quad {\parallel\delta\parallel }_{p} + c.f(x + \delta)\\
\textrm{such that} \quad x + \delta \in {[0,1]}^{n}
\IEEEyesnumber
\end{IEEEeqnarray*}

The algorithm will find $\delta$ such that it minimizes the distance metric, which is ${l}_{p}$ norm, and the objective function $f(x + \delta)$. $c$ is a constant to scale both the distance metric and objective function in the same range. Carlini and Wagner~\cite{carlini2017towards} used binary search to find the proper value for $c$. They explored several objective functions and found two that work the best. For a targeted attack scenarios with target class $t$, the best objective function is:

\begin{IEEEeqnarray*}{rCl}
f({x'}) = \max_{i \neq t}({F({x'})}_{i}) - {F({x'})}_{t}
\IEEEyesnumber
\end{IEEEeqnarray*}

For non-targeted attack scenarios where the true class for sample $x$ is class $y$, the best objective function is:
\begin{IEEEeqnarray*}{rCl}
f({x'}) =  {F(x')}_{y} - \max_{i \neq y}({F({x'})}_{i})
\IEEEyesnumber
\end{IEEEeqnarray*}

We consider two different attack scenarios based on when the defenses are applied. The defense can be applied either after the attack ({\em Without-Adversarial-Training})  or before the attack ({\em With-Adversarial-Training}). We use the same half-duplex (HD) closed-world dataset for these experiments mentioned in Section VI-A of the main paper. %We explain the pre-processing of the data in Section~\ref{dataset}.

\subsection{Without-Adversarial-Training}\label{scenarioI}
In this scenario, we assume that the attacker has trained its classifier on traces that have not been defended. We assume that the attacker is not aware of a defense in place. Such a scenario can be valid in the case that we have an attacker that does not target any specific client and his goal is to identify the behavior of large number of users, but some of the clients may use some defense to protect their traffic. %Here, the attacker is not aware of the defense or his goal is to monitor the majority of the clients.

For this scenario, we first train a classifier on our undefended traces and then we generate the adversarial traces. We examine the efficacy of the generated samples by testing them against the trained attacker model.

In our evaluation, we break the data into two sets, {\em Adv Set} and {\em Detector Set}. Each set has 83 classes with 360 instances each. The attacker trains the classifier on the Detector Set. 
%The traces in Detector Set are not protected by any defenses. 
The WF attacks that we apply on the Adv Set are the DF and CUMUL attacks. We chose DF as a state-of-the-art DL-based WF attack and CUMUL as a traditional ML-based WF attack. CUMUL uses an SVM model and has high performance compared to other traditional ML-based WF attacks~\cite{sirinamDF}.

%We apply the method described in~\ref{failed_methods} to generate adversarial traces from the traces in Adv Set, we call these traces in our evaluation as Adversarial Traces. 
To generate adversarial traces, we use a simple CNN as the detector, and the adversarial traces will be generated based on this simple CNN. The architecture of the simple CNN is shown in Table~\ref{cnn_modelarch}. 

%\vspace{-0.7cm}
%\input{tables/simple_CNN.tex}
\input{tables/undefended_traces.tex}
\input{tables/defended_traces.tex}

Table~\ref{attacker-first} shows the results of our evaluations. Adversarial traces add 62\% bandwidth overhead. When generated for the Simple CNN, the traces can confound the target model 98\% of the time. In addition, the accuracies of DF and CUMUL are 3\% and 31\%, respectively. This means that adversarial traces generated based on a target model with simple CNN architecture can be highly transferable to other ML models. Almost all the adversarial traces generated by simple CNN can confound DF attack, which is also a deep learning model. The results show that the adversarial traces are more transferable to the DL model than traditional ML models.

\subsection{With-Adversarial-Training}\label{scenarioII}

In this scenario, we assume that the attacker knows that the client is using some sort of defense mechanisms to protect her traffic. The attacker then collects the traces protected by the same method as the client and trains his classifier with those traces. In this scenario, the training set and testing set are both traces protected by the same WF defense method. This scenario is more realistic, since the attacker should have access to Tor's open-source code and could also just obtain a Tor client to directly generate defended traces.
%because it has been shown that the effectiveness of WF attacks depends on the attacker's knowledge of the clients~\cite{marc-css-2014}. Moreover, once a defense is deployed it is supposed to be accessible for all the users and used by all the users. Therefore, the attacker can also use  the same defense as other clients. 

For evaluation in this scenario, we generate adversarial traces using the C\&W method. Then we train the simple CNN, DF, and CUMUL attacks, on 90\% of the dataset of defended traces and test them with the remaining 10\% of the data. 

To generate adversarial traces, we train a target model with the same architecture as the simple CNN with the traces in the Detector Set and used in generating Adversarial Traces in the Adv Set. The results of the evaluation in this scenario are shown in Table~\ref{defender-first}. As shown in the table, even when adversarial traces are generated based on a target model with a similar architecture as the simple CNN, they are highly detectable on simple CNN trained on the adversarial traces, with an accuracy of 91\%. Moreover, DF and CUMUL attacks can also detect the adversarial traces with high accuracy, 97\% and 91\%, respectively. This means that generated adversarial traces are ineffective when the adversary adopts the technique of adversarial training. %Generating the adversarial traces works like a data augmentation technique in this case. If the attacker is trained on them, the attacker will detect them correctly. 
This highlights the necessity of the creation of a new adversarial example generation technique designed specifically for WF.

%% file: tables/undefended_traces.tex
\begin{table}[htp]
\centering

\caption{The evaluation of the defenses against the state-of-the-art WF attacks as the attackers are trained on the undefended traces. BWO: Bandwidth Overhead, CNN is the simple CNN of Table~\ref{cnn_modelarch}.}
\begin{tabular}{lcccc} 
\hline
    & BWO & CNN & DF~\cite{sirinamDF}   & CUMUL~\cite{cumul}  \\ 
\hline
Undefended         & -           & 92\%      & 98\% & 92\%   \\
Adversarial Traces & 62\%        & 2\%       & 3\%  & 31\%   \\
\hline
\end{tabular}
\label{attacker-first}
\end{table}

%% file: tables/defended_traces.tex
\begin{table}[!t]
\centering

\caption{The evaluation of the defenses against the state-of-the-art WF attacks as the attackers are trained on the defended traces. BWO: Bandwidth Overhead, CNN is the simple CNN of Table~\ref{cnn_modelarch}.}
\begin{tabular}{lcccc} 
\hline
                   & BWO & CNN & DF~\cite{sirinamDF}   & CUMUL~\cite{cumul}  \\ 
\hline
Undefended         & -           & 92\%      & 98\% & 92\%   \\
Adversarial Traces & 62\%        & 91\%       & 97\%  & 91\%   \\
\hline
\end{tabular}
\label{defender-first}
\end{table}

%% file: appen_D.tex
\section{\system~with C\&W}
\label{tunedcw}

\input{tables/simple_CNN.tex}

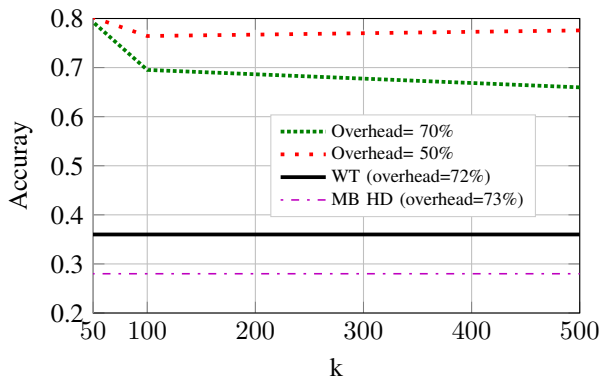
\begin{figure}[!t]
	\centering
	\input{figures/cw-mb.tex}
	\caption{DF attack accuracy on \system~with C\&W, {\em Mockingbird HD}, and W-T as the number of iterations $k$ varies.}
	\label{fig:cw-mb}
\end{figure}

Intuitively, there could be benefits from following an optimization methods such as that used in the Carlini and Wagner (C\&W) method~\cite{carlini2017towards} inside of the \system~algorithm. In theory, this could help the algorithm find an adversarial trace towards a selected target with lower perturbations and thus lower overhead while still benefiting from \system's random selection of targets. To evaluate this approach, we replaced the part of the \system~algorithm that moves the current trace towards a selected target trace with the C\&W algorithm that optimizes over the loss function to find the adversarial trace.

%We also mixed \system~with the . In this method, we use random target selection as in \system, but use C\&W to search for better adversarial examples. In this implementation, instead of gradient of distance, we optimized the adversarial traces over the loss function. 
Let's assume we have trace ${I}_{s}$ (belonging to source class $s$) and the generated adversarial trace is ${I}^{new}_{s}$: 

\begin{IEEEeqnarray*}{c}
{I}^{new}_{s} = {I}_{s} + \Delta
\IEEEyesnumber
\end{IEEEeqnarray*}

Where $\Delta$ the perturbation added to the source trace. We define $\Delta$ in a way that it does not add too much overhead:

\begin{IEEEeqnarray*}{c}
\Delta = \delta^{2} * scale
\IEEEyesnumber
\end{IEEEeqnarray*}

\noindent Here, $\delta$ is the perturbation optimized by the algorithm, and the $scale$ is a variable to adjust the overhead not to go above a maximum allowed overhead. Squaring is used to force perturbations to be non-negative. We define the constant variable $M$ as the maximum allowed overhead to the traces. $M$ should be between 0 and 1, which leads proportionally to between 0\% and 100\% bandwidth overhead.
%${I}_{s} = \left [ {b}^{I}_{0},  {b}^{I}_{1},...,{b}^{I}_{n}\right ]$

\begin{IEEEeqnarray*}{c}
 S = (1 +  M) * Size({I}_{s} )  - Size({I}^{new}_{s}) \\
 If \quad I = \left [ {b}^{I}_{0}, {b}^{I}_{1},...,{b}^{I}_{n}\right ]\quad then \\
Size(I) = \sum_{i=1}^{n} b_{i} 
 \IEEEyesnumber
\end{IEEEeqnarray*}
\begin{IEEEeqnarray*}{c}
scale = min\left(1 , \frac{S}{Size(\Delta)}\right)
\IEEEyesnumber
\end{IEEEeqnarray*}

Given a sample ${I}_{s}$ and a model $F$, the algorithm finds a perturbation $\delta$ that makes ${I}^{new}_{s} = {I}_{s} + \Delta$ to be misclassified to target class $t$. The algorithm tries to find ${I}^{new}_{s}$ that is similar to ${I}_{s}$. The algorithm will find $\delta$ such that it minimizes the objective function $f({I}^{new}_{s})$. C\&W~\cite{carlini2017towards} explored several objective functions and found that the best objective function is:
\begin{IEEEeqnarray*}{rCl}
f({{I}^{new}_{s}}) = \max_{i \neq t}({F({{I}^{new}_{s}})}_{i}) - {F({{I}^{new}_{s}})}_{t}
\IEEEyesnumber
\end{IEEEeqnarray*}

For each source trace ${I}_{s}$, we pick a target class $t$. We keep minimizing the objective function till ${I}^{new}_{s}$ is classified as class $t$.% or we reach the maximum of $k$ iterations. 
At the end of $k$ iterations, if ${I}^{new}_{s}$ is still not in class $t$, we change target $t$ to another class and minimize the objective function over the new target. We change the target a maximum of $T$ times. 

%We only show results for the half-duplex (HD) dataset. 

The architecture of our surrogate network is shown in Table~\ref{cnn_modelarch}. This surrogate network works as the {\em detector} (or model $F$). We trained the {\em detector} on {\em Detector Set} $\mathcal{D}$ and generate adversarial traces on {\em Adv Set} $\mathcal{A}$ on our half-duplex dataset.

In our experiments, we set $M$ equal to 0.5 and 0.7, which leads to 50\% and 70\% bandwidth overhead, respectively. We kept the maximum number of target changes fixed as $T=8$ and changed the maximum iteration of iterations $k$.  

Figure~\ref{fig:cw-mb} shows the accuracy of the {\em DF} attack on generated adversarial traces compared to W-T and {\em Mockingbird HD}. {\em Mockingbird HD} outperforms the generated adversarial traces with this tuned C\&W approach. With 70\% bandwidth overhead, this alternative approach of C\&W incurs 65\% attack accuracy which is at least 24\% higher than {\em Mockingbird HD}. With 50\% bandwidth overhead, the attack accuracy of this alternative approach is at least 35\% higher than any of the cases of {\em Mockingbird HD}.

This evaluation suggests that applying an optimization technique such as C\&W's method inside the {\em Mockingbird} algorithm is not as effective as simply reducing the distance between the current trace and a specific target trace. We believe that attempting to reduce the gradient leads to following a more predictable set of paths in the search space than following straight lines to unpredictable target samples. While it also may cost more overhead to find an adversarial example, the cost is worth paying to prevent adversarial training from being effective at finding the same kinds of adversarial examples.

%This evaluation shows not only that the base approach of C\&W is ineffective in the website fingerprinting domain but also a tuned approach remains ineffective.

%% file: tables/simple_CNN.tex
\begin{table}[t!]
\centering
\caption{Architecture of Simple CNN.}
\label{my-label}
\begin{tabular}{ll}
\hline
Layer type                & size                      \\ \hline
Convolution + ReLU        & 1 $\times$ 8 $\times$ 32  \\
Convolution + ReLU        & 1 $\times$ 8 $\times$ 64  \\
Convolution + ReLU        & 1 $\times$ 8 $\times$ 128 \\
Fully Connected + ReLU    & 512                       \\
Fully Connected + Softmax & number of classes    \\ \hline
\end{tabular}
\label{cnn_modelarch}
\vskip -0.3cm
\end{table}

%% file: figures/cw-mb.tex
\begin{tikzpicture}
\definecolor{color0}{rgb}{0.75,0,0.75}
\begin{axis}[
xlabel={k},
ylabel={Accuray},
xmin=50, xmax=500,
ymin=0.20, ymax=0.8,
xmajorgrids,
ymajorgrids,
width=1.150*\figurewidth,
height=1.10*\figureheight,
extra x ticks= {50},
ytick={0.2,0.3,0.4, 0.5, 0.6, 0.7, 0.8},
yticklabels={0.2,0.3,0.4, 0.5, 0.6, 0.7, 0.8},
legend entries={{Overhead= 70\%},{Overhead= 50\%},{WT (overhead=72\%)},{MB HD (overhead=73\%)}},
legend style={at={(0.91,0.5)}, anchor=east, draw=white!80.0!black, nodes={scale=0.7}},
legend cell align={left}
]
\addplot [ green!50.0!black, densely dotted, line width = 0.5mm]
table {%
500 0.65973851831
100 0.695273214904
50 0.792490781113
};
\addplot [line width = 0.5mm, red, loosely dotted]
table {%
500 0.775729131807
100 0.764331210211
50 0.802212537734
};
\addplot [line width = 0.5mm, black]
table {%
0 0.36
500 0.36
};
\addplot [semithick, color0, dash pattern=on 1pt off 3pt on 3pt off 3pt]
table {%
0 0.28
500 0.28
};
\end{axis}
\end{tikzpicture}